\RequirePackage{silence}
\documentclass[fleqn,usenatbib,useAMS]{mnras} 
\usepackage{graphicx}
\usepackage{amsmath}

\usepackage{amsfonts}
\usepackage{float}
\usepackage{bm}
\usepackage[perpage,symbol*]{footmisc}
\usepackage{ae,aecompl}
\usepackage{array}
\usepackage{soul}
\usepackage{mathtools}
\usepackage{multirow}

\newcommand{\appropto}{\mathrel{\vcenter{
		\offinterlineskip\halign{\hfil$##$\cr 
			\propto\cr\noalign{\kern2pt}\sim\cr\noalign{\kern-2pt}}}}}

\title{A new line on the wide binary test of gravity} 

\author[Indranil Banik]{Indranil Banik\thanks{Email:
\href{mailto:ibanik@astro.uni-bonn.de}{ibanik@astro.uni-bonn.de}}\\
Helmholtz-Institut f\"ur Strahlen und Kernphysik (HISKP), University of Bonn, Nussallee 14$-$16, D-53115 Bonn, Germany}

\pubyear{2019}
\begin{document}
\label{firstpage}
\pagerange{\pageref{firstpage}--\pageref{lastpage}}

\maketitle

\begin{abstract}

The relative velocity distribution of wide binary (WB) stars is sensitive to the law of gravity at the low accelerations typical of galactic outskirts. I consider the feasibility of this wide binary test using the `line velocity' method. This involves considering only the velocity components along the direction within the sky plane orthogonal to the systemic proper motion of each WB.

I apply this technique to the WB sample of Hernandez et. al. (2019), carefully accounting for large-angle effects at one order beyond leading. Based on Monte Carlo trials, the uncertainty in the one-dimensional velocity dispersion is $\approx 100$ m/s when using sky-projected relative velocities. Using line velocities reduces this to $\approx 30$ m/s because these are much less affected by distance uncertainties.

My analysis does not support the Hernandez et. al. (2019) claim of a clear departure from Newtonian dynamics beyond a radius of $\approx 10$ kAU, partly because I use $2\sigma$ outlier rejection to clean their sample first. Nonetheless, the uncertainties are small enough that existing WB data are nearly sufficient to distinguish Newtonian dynamics from Modified Newtonian Dynamics. I estimate that $\approx 1000$ WB systems will be required for this purpose if using only line velocities.

In addition to a larger sample, it will also be important to control for systematics like undetected companions and moving groups. This could be done statistically. The contamination can be minimized by considering a narrow theoretically motivated range of parameters and focusing on how different theories predict different proportions of WBs in this region.

\end{abstract}


\section{Introduction}
\label{Introduction}

The standard cosmological paradigm relies on the assumption that General Relativity applies accurately to all astronomical systems. On Solar System and galaxy scales, it can be well approximated by Newtonian gravity due to the non-relativistic speeds \citep{Rowland_2015, Almeida_2016}. Newtonian gravity was originally designed to explain motions inside the Solar System. However, it appears to break down when extrapolated to galaxies, where the gravitational field can often be estimated from rotation curves \citep[e.g.][]{Babcock_1939, Rubin_Ford_1970, Rogstad_1972}. These acceleration discrepancies are thought to be caused by halos of cold dark matter (CDM) surrounding each galaxy \citep{Ostriker_Peebles_1973}.

Unfortunately, the nature of this CDM remains elusive. Gravitational microlensing experiments indicate that the Galactic CDM can't be made of compact objects like stellar remnants \citep{MACHO_2000, EROS_2007}. The microlensing timescale would become longer than the survey duration if the CDM was made of heavier objects like primordial black holes \citep[e.g.][]{Carr_2016, Clesse_2018}. However, this idea runs into difficulties when confronted with data on gravitational lensing of quasars \citep{Mediavilla_2017} and supernovae \citep{Zumalacarregui_2018}.

Thus, CDM is generally considered to be an undiscovered weakly interacting particle beyond the well-tested standard model of particle physics \citep[][and references therein]{Peebles_2017_DM_review}. Despite extensive searches for such particles, none have so far been detected, ruling out a large part of the available parameter space \citep{Liu_2017}. Searches for the effects of dynamical friction on the extensive CDM halos have also turned up empty-handed \citep{Angus_2011, Kroupa_2015, Oehm_2017, Oehm_2018}. Major tensions between observations and theory also indicate that a revision of CDM-based models may be needed \citep{Kroupa_2012}, in particular due to the anisotropic distribution of Local Group satellite galaxies \citep{Pawlowski_2018}.

Given these results, it is prudent to question the underlying assumption of Newtonian gravity \citep{Zwicky_1937} and its general relativistic extension. This has proved very successful in a wide variety of tests, most famously when it correctly predicted how much light would be deflected by the Solar gravitational field \citep{Eddington_1919}. More recently, it has passed tests based on the gravitational redshift of objects near the Galactic Centre black hole \citep{Gravity_2018, Gravity_2019}.

However, these successes do not prove that the theory can be extrapolated to the much larger scale of galaxies, where accelerations are typically much less than in the Solar System. In fact, a failure of this extrapolation can naturally explain the remarkably tight correlation between the internal accelerations within galaxies and the predictions of Newtonian gravity applied to their luminous matter distributions \citep[e.g.][and references therein]{Famaey_McGaugh_2012}. This `radial acceleration relation' (RAR) has recently been tightened significantly with near-infrared photometry from the Spitzer Space Telescope \citep{SPARC}, considering only the most reliable rotation curves (see their section 3.2.2) and exploiting reduced variability in stellar mass-to-light ratios at near-infrared wavelengths \citep{Bell_de_Jong_2001, Norris_2016}. These improvements show that the RAR holds with very little scatter over ${\approx 5}$ orders of magnitude in luminosity and a similar range in surface brightness \citep{McGaugh_Lelli_2016}. Fits to individual rotation curves show that the intrinsic scatter in the RAR must be ${<13\%}$ and is consistent with 0 \citep{Li_2018}. Although \citet{Rodrigues_2018} claimed that some rotation curves do not satisfy the RAR, it was later shown that these cases mostly arise when the distance is particularly uncertain \citep{Kroupa_2018}. For galaxies where this is known well, discrepancies with the RAR are rather mild and may be caused by small yet unmodelled effects like disk warping and gradients in the stellar mass-to-light ratio.

These recent observations were predicted several decades earlier by the theory called Modified Newtonian Dynamics \citep[MOND,][]{Milgrom_1983}. In MOND, the dynamical effects usually attributed to CDM are instead provided by an acceleration-dependent modification to gravity, similar to the behaviour which arises in the mimetic model \citep{Vagnozzi_2017}. The basic idea is that the gravitational acceleration $g$ at distance $r$ from an isolated point mass $M$ transitions from the Newtonian $\frac{GM}{r^2}$ law at short range to
\begin{eqnarray}
	g ~=~ \frac{\sqrt{GMa_{_0}}}{r} ~~~\text{for } ~ r \gg \overbrace{\sqrt{\frac{GM}{a_{_0}}}}^{r_{_M}} \, .
	\label{Deep_MOND_limit}
\end{eqnarray}

MOND (or Milgromian dynamics) introduces $a_{_0}$ as a fundamental acceleration scale of nature below which the deviation from Newtonian dynamics becomes significant and the equations of motion become spacetime scale invariant \citep{Milgrom_2009_DML}. Empirically, $a_{_0} \approx 1.2 \times {10}^{-10}$ m/s$^2$ to match galaxy rotation curves \citep{McGaugh_2011}. Remarkably, this is within an order of magnitude of the acceleration at which the classical energy density in a gravitational field \citep[][equation 9]{Peters_1981} becomes comparable to the dark energy density $u_{_\Lambda} \equiv \rho_{_\Lambda} c^2$ that conventionally explains the accelerating expansion of the Universe \citep{Ostriker_Steinhardt_1995, Riess_1998, Perlmutter_1999}.
\begin{eqnarray}
	\frac{g^2}{8\mathrm{\pi}G} ~<~ u_{_\Lambda} ~~\Leftrightarrow~~ g ~\la~ 2\mathrm{\pi}a_{_0} \, .
	\label{MOND_quantum_link}
\end{eqnarray}

MOND could thus be a result of poorly understood quantum gravity effects \citep[e.g.][]{Milgrom_1999, Pazy_2013, Verlinde_2016, Smolin_2017}. Regardless of its underlying microphysical explanation, it can accurately match the rotation curves of a wide variety of both spiral and elliptical galaxies across a vast range in mass, surface brightness and gas fraction \citep[][and references therein]{Lelli_2017}. MOND does all this based solely on the distribution of luminous matter. Given that most of these rotation curves were obtained in the decades after the MOND field equation was first published \citep{Bekenstein_Milgrom_1984}, these achievements are successful a priori predictions. These predictions work due to regularities in rotation curves that are difficult to reconcile with collisionless halos of CDM whose nature is very different to baryons \citep{Salucci_2017, Desmond_2016, Desmond_2017}.

Because MOND is an acceleration-dependent theory, its effects could become apparent in a rather small system if this has a sufficiently low mass. In fact, the MOND radius $r_{_M}$ is only 7000 astronomical units (7 kAU) for a system with $M = M_\odot$ (Equation \ref{Deep_MOND_limit}). This suggests that the orbits of distant Solar System objects might be affected by MOND \citep{Pauco_2016}, possibly accounting for certain correlations in their properties \citep{Pauco_2017}. However, it is difficult to accurately constrain the dynamics of objects at such large distances.

Such constraints could be obtained more easily around other stars if they have distant binary companions. As first suggested by \citet{Hernandez_2012}, the orbital motions of these wide binaries (WBs) should be faster in MOND than in Newtonian gravity. Moreover, it is likely that many such systems would form \citep{Kouwenhoven_2010, Tokovinin_2017}, paving the way for the wide binary test (WBT) of gravity that I discuss in this contribution.

WBs are likely to comprise at least a few percent of stellar systems given that the nearest star to the Sun is in a WB. Proxima Centauri orbits the close (18 AU) binary $\alpha$ Centauri A and B at a distance of 13 kAU \citep{Kervella_2017}. The Proxima Centauri orbit would thus be significantly affected by MOND \citep{Beech_2009, Beech_2011}, a prediction that might be directly testable with next-generation astrometry \citep{Banik_2019_Proxima}. Given the billions of stars in our Galaxy, it almost certainly contains a vast number of systems well suited to the WBT. This is especially true given the high (74\%) likelihood that our nearest WB was stable over the last 5 Gyr despite the effects of Galactic tides and stellar encounters \citep{Feng_2018}. This system was probably also stable in MOND \citep[][section 9]{Banik_2018_Centauri}.

Proxima Centauri is far from the only WB within reach of existing observations. Data from the Gaia mission \citep{Perryman_2001} strongly suggests the presence of several thousand WBs within $\approx 150$ pc \citep{Andrews_2017}. The candidate systems they identified are mostly genuine, with a contamination rate of ${\approx 6\%}$ \citep{Andrews_2018} estimated using the second data release of the Gaia mission \citep[Gaia DR2,][]{GAIA_2018}.

The WBT was considered in more detail by \citet{Pittordis_2018}, who set up simulations of WBs in Newtonian gravity and several theories of modified gravity, including MOND. These calculations were revisited by \citet{Banik_2018_Centauri} using self-consistent MOND simulations that include the external field from the rest of the Galaxy (Section \ref{No_EFE}) and use an interpolating function consistent with the RAR. Their main result was that MOND enhances the orbital velocities of Solar neighbourhood WBs by ${\approx 20\%}$ above Newtonian expectations, consistent with their analytic estimate (see their section 2.2). Using statistical methods they developed, they showed that $\approx 500$ WB systems would be required to detect this effect if measurement errors are neglected but only the more accurately known sky-projected quantities are used. They also considered various systematic errors which could hamper the WBT, in particular the presence of a low mass undetected companion to one of the stars in a WB (see their section 8.2).

The WBT was first attempted by \citet{Hernandez_2012} using the WB catalogue of \citet{Shaya_2011}, who analyzed Hipparcos data with Bayesian methods to identify WBs within 100 pc \citep{Leeuwen_2007}. Typical relative velocities $\bm{v}_{rel}$ between WB stars seemed to remain constant with increasing separation instead of following the expected Keplerian decline \citep[][figure 1]{Hernandez_2012}. However, it was later shown that their typical velocity uncertainty of 800 m/s was too large to draw strong conclusions about the underlying law of gravity \citep[][section 1]{Scarpa_2017}. This is because the typical velocity scale of the WBT is $\sqrt[4]{GM_\odot a_{_0}} = 360$ m/s (Equation \ref{Deep_MOND_limit} at $r = r_{_M}$).

Recently, \citet{HERNANDEZ_2018} revisited their earlier WB sample using Gaia DR2, focusing on only sky-projected relative velocities due to the time required to obtain follow-up spectroscopic redshift measurements and difficulties in correcting these for stellar convective blueshifts \citep[][section 2.2]{Kervella_2017}. Unfortunately, the \citet{HERNANDEZ_2018} analysis suffers from a deficiency related to incorrect visualisation of how spherical co-ordinate systems work \citep{Badry_2019}. These perspective effects were discussed in more detail by \citet[][section 3.2]{Shaya_2011} and \citet[][section 2.4]{Pittordis_2018}.

Such effects can broadly be understood by considering a WB composed of stars $A$ and $B$ in sky directions $\widehat{\bm{n}}_A$ and $\widehat{\bm{n}}_B$, respectively. If we are interested in their sky-projected relative velocity $\bm{v}_{sky}$ and define this as that part of $\bm{v}_{rel}$ within the plane orthogonal to $\widehat{\bm{n}}_A$, then only the proper motion of star $A$ is required. However, for star $B$, we also need to know its radial velocity because $\widehat{\bm{n}}_B \neq \widehat{\bm{n}}_A$, causing $\widehat{\bm{n}}_B$ to partly lie within the plane orthogonal to $\widehat{\bm{n}}_A$. In general, knowledge of both stars' radial velocities is required under other definitions of the sky plane e.g. Equation \ref{n_hat_sys}. However, the analysis of \citet{HERNANDEZ_2018} did not consider radial velocity information, implicitly assuming that $\widehat{\bm{n}}_B = \widehat{\bm{n}}_A$ \citep{Badry_2019}.\footnote{While this article was under review, \citet{HERNANDEZ_2018} corrected this deficiency in a subsequent reanalysis. Their overall conclusions remain very similar.}

As well as correcting this deficiency, I consider how to reduce the uncertainty in $\bm{v}_{rel}$. Part of this is due to uncertainty in the relative heliocentric distances to the stars in a WB, which can be difficult to constrain (Section \ref{Discussion_tangential_velocity}). To quantify the effect this has, suppose that the typical heliocentric tangential velocity of the system is 30 km/s \citep{GAIA_2018}. With a 1\% distance uncertainty, even a perfectly measured proper motion implies a velocity uncertainty of ${\approx 300}$ m/s. This is nearly the same as WB relative velocities of $\sim 300$ m/s \citep[][figure 7]{Banik_2018_Centauri}. Thus, if $\bm{v}_{rel}$ is parallel to the WB systemic proper motion, even rather small distance uncertainties would make it very challenging to accurately infer $\bm{v}_{rel}$. This is a serious limitation on the WBT because distances are expected to be less accurately known than proper motions (Section \ref{Discussion_tangential_velocity}). For example, Gaia DR2 parallaxes have a zero-point offset which probably varies with magnitude and colour \citep{GAIA_2018, Riess_2018}. This might seriously complicate the WBT or weaken its statistical significance by restricting it to only those WBs that consist of similar stars.

To a large extent, the distance issue can be avoided if $\bm{v}_{rel}$ is orthogonal to the WB systemic proper motion (Equation \ref{v_tan_determination}). I consider the feasibility of exploiting this using the `line velocity method' for the WBT. The idea is to use only one component of $\bm{v}_{rel}$, namely that within the sky plane and orthogonal to the systemic proper motion of the WB \citep[][section 3.2]{Shaya_2011}. Because the WBT is statistical in any case, it is possible using line velocities in a similar way to if using both components of the sky-projected velocity. In the following, when discussing use of the sky-projected velocity, I mean both components thereof.

Line velocities minimize the impact of uncertainty in the relative distance. However, it is also possible to better constrain this using the accurately known projected separation of a WB (Section \ref{rp_trick}). This technique has the advantage of using two components of $\bm{v}_{rel}$, which could significantly improve the statistical power of the WBT.

After explaining these methods more precisely in Section \ref{Line_velocity_method}, I apply them to the \citet{HERNANDEZ_2018} dataset to confirm that both significantly reduce uncertainties compared to the use of conventional sky-plane velocities (Section \ref{Application_to_Hernandez}). With line velocities, the main disadvantage is that the use of essentially half as much data from each WB roughly doubles the number of systems required for the WBT if measurement uncertainties are neglected (Section \ref{Effect_on_P_detection}). I discuss future prospects for the WBT in Section \ref{Discussion}, where I explain why the line velocity method is likely to prove very fruitful in the long run. My conclusions are given in Section \ref{Conclusions}.

\section{Quantifying the relative velocity}
\label{Line_velocity_method}

The basic idea behind the line velocity method is to focus on the directions rather than the magnitudes of the proper motion vectors of the stars in a WB. This is because converting a difference in proper motion directions into a relative velocity only requires the distance to the WB system as a whole. However, a difference in proper motion magnitudes can only be converted into a relative velocity if observers also know the \emph{relative} distances to its stars. Because distances are likely to be less accurately known than proper motions (Section \ref{Discussion_tangential_velocity}), the line velocity method uses only the most reliably known component of $\bm{v}_{rel}$. In this section, I explain how to apply this method and a related method involving additional assumptions on the relative distance (Section \ref{rp_trick}).

\subsection{The sky-projected separation}
\label{Sky_projected_separation}

The WBT is ideally performed using accurate 3-dimensional (3D) positions and velocities. Unfortunately, Gaia DR2 distance uncertainties are ${\approx 80}$ kAU for a system 100 pc from the observatory \citep[][section 6.2]{Banik_2018_Centauri}. This is much larger than the $3-20$ kAU range of separations recommended by that work for the WBT. Even if a slightly larger range is used, it is clear that observers do not reliably know the true 3D separation $r_{rel}$ for the vast majority of WBs. An exception arises for very nearby systems like $\alpha$ Centauri \citep{Kervella_2016}, but I expect there will be too few such systems to enable the WBT. Somehow, more distant systems must also be considered.

Fortunately, these systems can be utilized in a statistical sense if one uses their accurately known sky-projected separation $r_{sky}$ \citep{Pittordis_2018}. Thus, the WBT in the short term will be based on $r_{sky}$ and $\bm{v}_{rel}$. It is possible that not all components of $\bm{v}_{rel}$ will be used as they are not all equally well measured. This is the main issue I consider in this contribution.

To understand how $\bm{v}_{rel}$ can be obtained from the observables of a WB, I define unit vectors $\widehat{\bm{n}}_1$ and $\widehat{\bm{n}}_2$ towards each of its stars. I use the convention that any vector $\bm{v}$ has length $v \equiv \left| \bm{v} \right|$ such that the unit vector parallel to $\bm{v}$ is $\widehat{\bm{v}} \equiv \bm{v} \div v$. Given the observed heliocentric distance $d_i$ of star $i$, its position is thus
\begin{eqnarray}
    \bm{r}_{i} ~\equiv~ d_i \widehat{\bm{n}}_i ~~ \left(i = 1, \, 2 \right) \, .
\end{eqnarray}

The separation vector between the stars is $\bm{r}_{rel} \equiv \bm{r}_2 - \bm{r}_1$ and their relative velocity is $\bm{v}_{rel} \equiv \bm{v}_2 - \bm{v}_1$. The velocity $\bm{v}_{i}$ is found for each star individually from its radial velocity, proper motion, distance and sky position.

To define $r_{sky}$, I need to estimate the line of sight $\widehat{\bm{n}}_{sys}$ towards the system as a whole. Given the small angular separation of a WB, I assume $\widehat{\bm{n}}_{sys}$ is directed towards the mid-point of its stars.
\begin{eqnarray}
\widehat{\bm{n}}_{sys} ~\propto~ \frac{d_1 \widehat{\bm{n}}_1 + d_2 \widehat{\bm{n}}_2}{2} \, .
\label{n_hat_sys}
\end{eqnarray}
Their sky-projected separation is thus
\begin{eqnarray}
r_{sky} ~\equiv~ \left| \bm{r}_{rel} - \left( \bm{r}_{rel} \cdot \widehat{\bm{n}}_{sys} \right) \widehat{\bm{n}}_{sys} \right| \, .
\label{r_p_determination}
\end{eqnarray}

\subsection{The relative velocity}
\label{Relative_velocity}

I calculate the sky-projected relative velocity $v_{sky}$ analogously to Equation \ref{r_p_determination}.
\begin{eqnarray}
v_{sky} ~\equiv~ \left| \bm{v}_{rel} - \left( \bm{v}_{rel} \cdot \widehat{\bm{n}}_{sys} \right) \widehat{\bm{n}}_{sys} \right| \, .
\label{v_p_determination}
\end{eqnarray}

To apply the line velocity method, I need to estimate the systemic motion $\bm{v}_{sys}$ of the WB. Given that this is typically much larger than $\bm{v}_{rel}$, I determine $\bm{v}_{sys}$ under the simplifying assumption of an equal mass WB.
\begin{eqnarray}
\bm{v}_{sys} ~=~ \frac{\bm{v}_1 + \bm{v}_2}{2} \, .
\label{v_sys}
\end{eqnarray}


The line velocity method involves finding the component of $\bm{v}_{rel}$ along the direction $\widehat{\bm{v}}_{_{line}}$, the line within the sky plane orthogonal to $\bm{v}_{sys}$.
\begin{eqnarray}
\widehat{\bm{v}}_{_{line}} ~\propto~ \widehat{\bm{n}}_{sys} \times \widehat{\bm{v}}_{sys} \, .
\label{v_line_direction}
\end{eqnarray}

It is also possible to think of $\widehat{\bm{v}}_{_{line}}$ as the direction within the sky plane orthogonal to the systemic tangential velocity $\bm{v}_{tan}$ of the WB, where
\begin{eqnarray}
\bm{v}_{tan} ~\equiv~ \bm{v}_{sys} - \left( \bm{v}_{sys} \cdot \widehat{\bm{n}}_{sys} \right) \widehat{\bm{n}}_{sys} \, .
\label{v_tan_determination}
\end{eqnarray}

Having determined $\widehat{\bm{v}}_{_{line}}$, it is simple to determine the relative velocity of the stars along this line.
\begin{eqnarray}
v_{_{line}} ~\equiv~ \left| \bm{v}_{rel} \cdot \widehat{\bm{v}}_{_{line}} \right| \, .
\label{v_line}
\end{eqnarray}

In the rest of this work, I use three different measures of relative velocity corresponding to using 1, 2 or 3 of its components in the co-ordinate system defined by the orthogonal vectors $\widehat{\bm{v}}_{_{line}}$ and $\widehat{\bm{n}}_{sys}$. The simplest case is when using the full $\bm{v}_{rel}$. The 2D case corresponds to using $v_{sky}$ while the 1D case involves $v_{_{line}}$ alone.

\subsection{Estimating relative distances from sky positions}
\label{rp_trick}

For reasons I discuss in Section \ref{Discussion_tangential_velocity}, $v_{sky}$ can suffer from considerable uncertainty due to that in the relative distances to the stars in a WB. If this is estimated directly from Gaia DR2, the uncertainty is likely to be ${\approx 80 \sqrt{2}}$ kAU at a typical distance of 100 pc \citep[][section 6.2]{Banik_2018_Centauri}. This vastly exceeds the maximum $r_{sky}$ of systems used for the WBT (e.g. that work suggested using a maximum of 20 kAU).

Statistically, one expects that $r_{sky}$ is comparable to the line of sight separation $r_{rel, LOS} \equiv d_2 - d_1$. Thus, the uncertainty in $r_{rel, LOS}$ should not be much larger than $r_{sky}$ itself. To exploit this, I apply Bayes' Theorem to infer the total 3D separation $r_{rel}$ from the accurately known $r_{sky}$.

For a given $r_{rel}$, the conditional probability of observing a particular $r_{sky}$ can be found using basic spherical geometry. Defining a temporary variable $R_{sky}$, I get that
\begin{eqnarray}
	P \left( r_{sky} \left| r_{rel} \right. \right) ~\equiv~ \frac{\partial}{\partial r_{sky}} \left[ P \left( R_{sky} \leq r_{sky} \left| r_{rel} \right. \right)\right] \, .
\end{eqnarray}
This can be simplified by defining an angle $\theta = \sin^{-1} \left( r_{sky}/r_{rel} \right)$ between the line of sight and 3D separation vectors.
\begin{eqnarray}
	P \left( r_{sky} \left| r_{rel} \right. \right) &=& \frac{\partial}{\partial r_{sky}} \left( 1 - \cos \theta \right) \\
	&=& \frac{\sin \theta}{r_{rel} \cos \theta} \\
	&=& \frac{r_{sky}}{r_{rel} \sqrt{{r_{rel}}^2 - {r_{sky}}^2}} \\
	&=& \frac{1}{r_{sky} \, x \sqrt{x^2 - 1}} \, \text{, where} \\
	x &\equiv& \frac{r_{rel}}{r_{sky}} \geq 1 \, .
\end{eqnarray}

Bayes' Theorem also requires a prior on $r_{rel}$, or equivalently on $x$. For this, I use the WB results of \citet{Andrews_2017}, who found that $r_{sky}$ follows a power law with slope $-\alpha = -1.6$. The WB semi-major axes can be expected to follow a similar distribution, suggesting that the same is true for $r_{rel}$ \citep[][section 3.2]{Banik_2018_Centauri}. A more refined prior could be used for the WBT. For the moment, I simply assume a prior on $x$ of
\begin{eqnarray}
	P_{prior} \left( x \right) ~\propto~ x^{-\alpha} \, ,
\end{eqnarray}
implying a posterior of
\begin{eqnarray}
	P \left( x \right) ~\propto~ \frac{x^{-\left( \alpha + 1 \right)}}{\sqrt{x^2 - 1}} \, , \, x \geq 1.
	\label{x_posterior}
\end{eqnarray}
I obtain the total normalisation by numerically integrating this, though I use analytic approximations when $x \approx 1$ and $x \gg 1$ to avoid numerical issues. I then obtain the cumulative distribution function of $x$ by numerically integrating its distribution $P \left( x \right)$, again reverting to analytic approaches for $x \approx 1$ and $x \gg 1$. This allows $x$ to be sampled from its distribution (Section \ref{Measurement_uncertainties}).

\section{Application to the \citet{HERNANDEZ_2018} sample}
\label{Application_to_Hernandez}

To see how the methods outlined in Section \ref{Line_velocity_method} work with real data, I apply them to the WB sample of \citet[][table 2]{HERNANDEZ_2018}, taking only their sky positions, radial velocities, distances and proper motions since I calculate all derived quantities myself.\footnote{Machine-readable versions are available in \textsc{excel}$^\text{\textregistered}$ and text formats upon request to the author. Note that the $\widehat{\bm{n}}_{i}$ are given in the International Celestial Reference System \citep{Ma_1998}.} Their work is based on using Gaia DR2 \citep{GAIA_2018} to update the astrometry of the carefully selected WB sample in \citet{Shaya_2011}, who used data from the Hipparcos mission \citep{Perryman_1997}.

\subsection{Quality cuts}
\label{Quality_cuts}

As might be expected, not all systems in the \citet{HERNANDEZ_2018} catalogue are usable in my analysis. Its table 2 has a column labelled `Exclusion Test' specifically for the purpose of flagging systems with a serious observational inconsistency or a large change in velocity between the Hipparcos and Gaia epochs, suggestive of another component in the system \citep[][section 8.2]{ Banik_2018_Centauri}. In this work, I only consider systems where the `Exclusion Test' column has a blank entry for both stars, indicating that \citet{HERNANDEZ_2018} found no good reason to reject the WB from their analysis.

As explained in Section \ref{Introduction} and pointed out by \citet{Badry_2019}, determining even just the sky-projected relative velocity requires radial velocity measurements. Thus, I reject systems where these are not available for either star. If it is known for one star, then I assume the same value and uncertainty for the other star and use the system in my analysis. This is because the difference in radial velocity is one component of $\bm{v}_{rel}$ and thus likely to be ${\approx 300}$ m/s for a genuine WB \citep[][figure 7]{Banik_2018_Centauri}. The contribution of this to $v_{sky}$ is smaller by at least a factor of the sky angle between the WB's components. Even a WB with $r_{sky} = 50$ kAU located 20 pc away subtends an angle of only 0.012 radians on our sky, implying that radial velocities affect $v_{sky}$ by ${\la 5}$ m/s. Thus, the small sky angles involved mean that my results should not be much affected by ignoring the \emph{difference} in radial velocities. Moreover, my main objective in this work is to quantify the typical uncertainty on $v_{sky}$ and $v_{_{line}}$ (Section \ref{Measurement_uncertainties}). The radial velocity should play only a small part in this, as long as that of the system is known reasonably well. In Section \ref{Discussion_radial_velocity}, I discuss some possible approaches in case this is not known.

\begin{table}
	\centering
	\begin{tabular}{cccc}
		\hline
		& & \multicolumn{2}{c}{Number of systems} \\
		Bin & Range in $r_{sky}$, kAU & before & after \\
		& & \multicolumn{2}{c}{rejection of outliers} \\
		\hline
		1 & 0.69 $-$ 3.27 & 21 & 20 \\
		2 & 3.27 $-$ 16.38 & 24 & 22 \\
		3 & 16.38 $-$ 82.12 & 17 & 9 \\
		4 & 82.12 $-$ 411.55 & 7 & 5 \\
		5 & 411.55 $-$ 1766.29 & 10 & 9 \\
		\hline
	\end{tabular}
	\caption{How I bin the data of \citet{HERNANDEZ_2018} in $r_{sky}$ (Equation \ref{r_p_determination}). Each bin is $\approx 0.7$ dex wide, similar to the bins they used. The numbers in the last two columns indicate how many systems remain in each bin before and after 2$\sigma$ outlier rejection (Section \ref{Quality_cuts}).}
	\label{r_sky_edges}
\end{table}

Upon examination of the systems which pass my selection criteria, it is evident that systems 822 and 823 in the \citet{Shaya_2011} catalogue share a star, making this a triple system.\footnote{I was advised of this by R. A. M. Cort{\'e}s.} Although this is a rather hierarchical system, I reject it from my analysis due to the additional complications which might nonetheless arise in a non-linear gravity theory (Section \ref{No_EFE}) and in three-body systems more generally.

For my analysis, I bin the remaining 79 systems in $r_{sky}$. The bins correspond as closely as possible to those used by \citet{HERNANDEZ_2018}, who used bins of width 0.7 dex. The bins used in this work are listed in Table \ref{r_sky_edges}.

I then apply my line velocity method to determine the root mean square (rms) $v_{_{line}}$ of the systems in each bin. My results in Section \ref{Effect_on_P_detection} show that their line velocities should follow a roughly Gaussian distribution. Thus, I apply a basic outlier rejection system to remove WBs whose $v_{_{line}}$ exceeds twice the rms value for systems in the same $r_{sky}$ bin. This reduces the estimated one-dimensional velocity dispersion $\sigma_{_{1D}}$, so the process is continued iteratively until it converges and no more WBs are rejected. In this way, I am left with 65 systems for the rest of my analysis. Their $r_{sky}$ distribution is shown in the final column of Table \ref{r_sky_edges}.

\subsection{Measurement uncertainties}
\label{Measurement_uncertainties}

To better estimate the uncertainty on $\sigma_{_{1D}}$, I conduct a control analysis in which I set $\bm{v} = \bm{v}_{sys}$ (Equation \ref{v_sys}) for both stars in a WB. The idea is to determine the rms relative velocity in different $r_{sky}$ bins \emph{if no actual velocity dispersion exists}. Non-zero values arise entirely from uncertainties in distances, radial velocities and proper motions. To account for these uncertainties, I perform $10^6$ Monte Carlo (MC) trials where I vary these quantities randomly according to their measurement errors, which I take to follow independent Gaussian distributions.\footnote{Because the distances to the stars I analyze are generally known to within 1\%, a Gaussian error on the parallax very nearly translates into a Gaussian distance distribution \citep{Luri_2018}.} Each time, I recompute $v_{rel}$, $v_{sky}$ and $v_{_{line}}$. To speed up the computations, I make use of the fact that changes in the distance or proper motion have a linear effect on the velocity. I include the cross-term that arises because in general both the distance and proper motion differ from their observed values and these must be multiplied to obtain a velocity.

\begin{figure}
	\centering
	\includegraphics[width = 8.5cm] {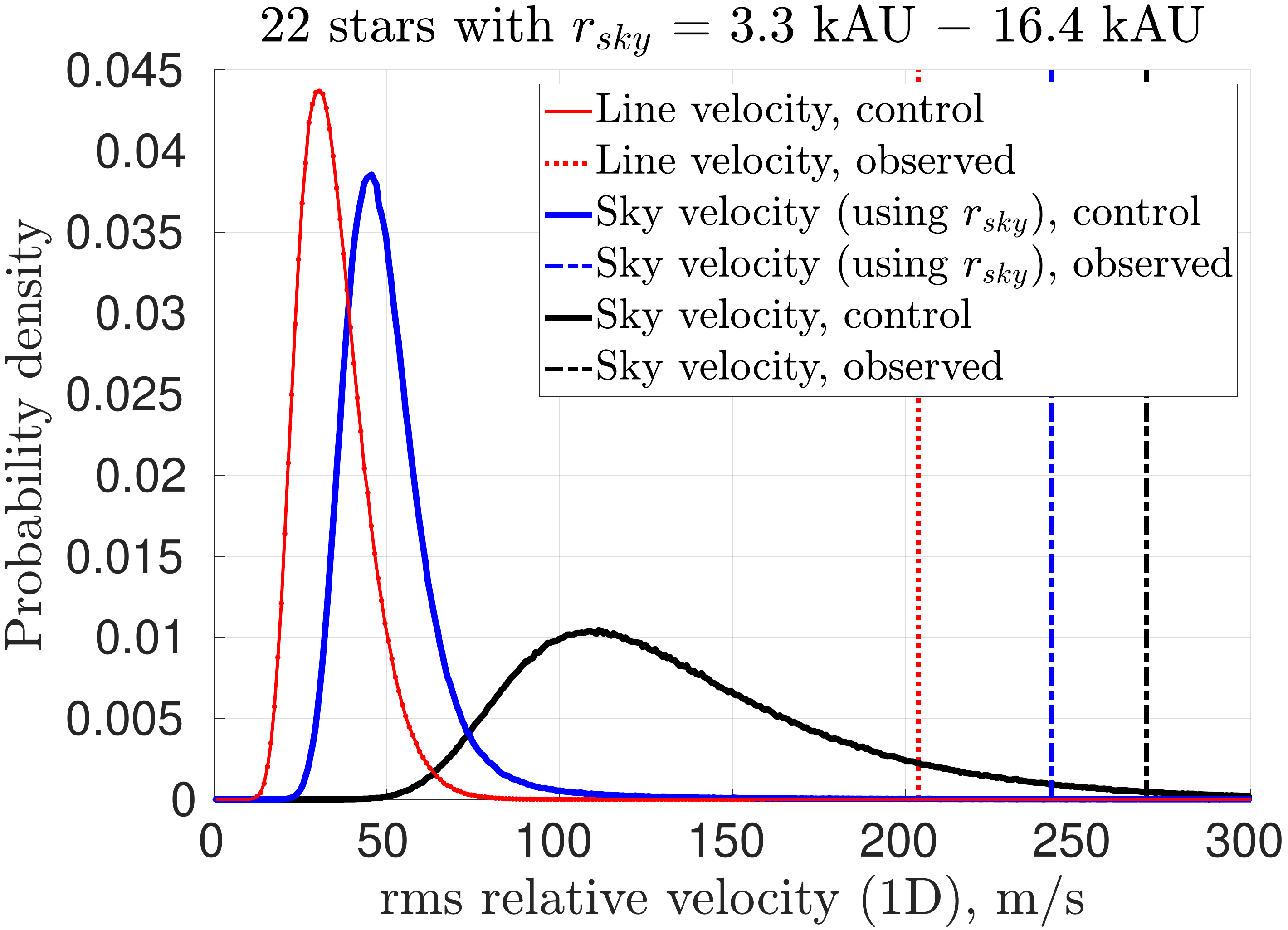}
	\caption{The distribution of the rms $v_{_{line}}$ (red) in my control Monte Carlo analysis of the WBs in bin 2 (Table \ref{r_sky_edges}). As the WBs are assumed to have zero relative velocity, non-zero values arise entirely from measurement uncertainties. I also show results obtained using $v_{sky}$, calculated both conventionally (black; Equation \ref{v_p_determination}) and using the method of Section \ref{rp_trick} where this is likely to yield better results (blue; see text). The observed values are shown using non-solid vertical lines with the same colour. Simulated and observed velocities based on $v_{sky}$ are scaled down by $\sqrt{2}$ to allow a fair comparison. Notice how the observed rms $v_{sky}$ is consistent with zero relative velocity in all 22 systems if it is calculated conventionally. This is not true for $v_{_{line}}$ or for $v_{sky}$ if the latter is calculated using the method of Section \ref{rp_trick}.}
	\label{Bin_2_v_line_sky}
\end{figure}

Measurement uncertainties influence the stellar velocities and thus the systemic velocity (Equation \ref{v_sys}), slightly affecting the direction of $\widehat{\bm{v}}_{_{line}}$ (Equation \ref{v_line_direction}). Because $v_{rel} \ll v_{sys}$, I use a small angle approximation to estimate how much $\widehat{\bm{v}}_{_{line}}$ should be rotated within the plane orthogonal to $\widehat{\bm{n}}_{sys}$, which I assume is unaffected by changes in the individual $d_i$ (Equation \ref{n_hat_sys}).

\begin{figure}
	\centering
	\includegraphics[width = 8.5cm] {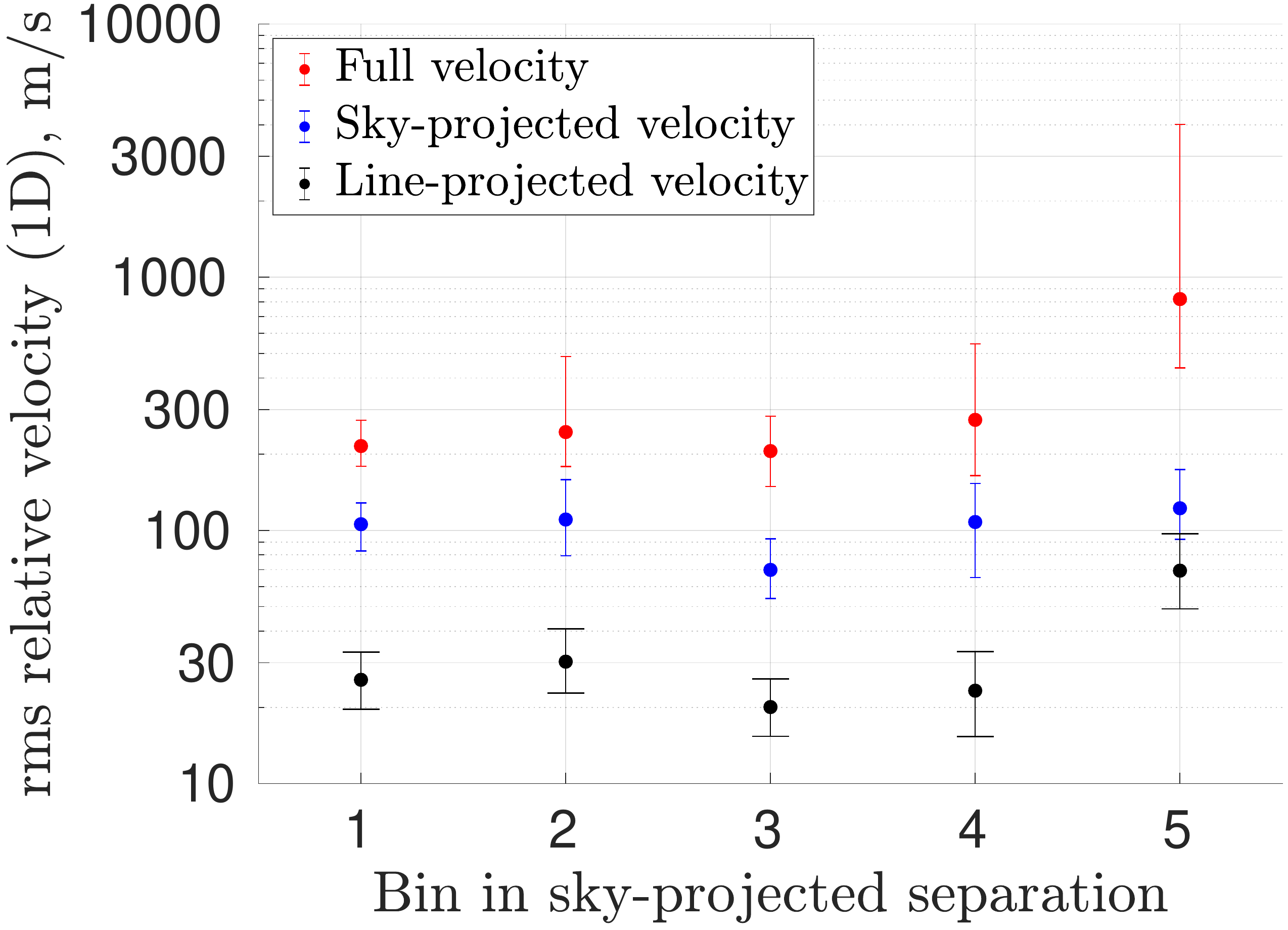}
	\caption{Similar to Figure \ref{Bin_2_v_line_sky}, but now showing results for all $r_{sky}$ bins (Table \ref{r_sky_edges}) and for the case where the full 3D relative velocity is considered (red points). The method of Section \ref{rp_trick} is not applied here. All velocities shown are scaled down by the square root of the number of dimensions to allow a fair comparison. Each probability distribution is summarised by its mode and 68.3\% confidence interval (see text). For systems where only one star has a measured radial velocity, I assume the same value and uncertainty for the other star before averaging the resulting 3D velocities and assigning the mean to both stars.}
	\label{Hernandez_v_control}
\end{figure}

When applying the method outlined in Section \ref{rp_trick}, I use a similar procedure but with some important differences. Since this technique constrains $r_{rel, LOS}$ to within ${\approx r_{sky}}$, I first check whether the quadrature sum of the Gaia distance uncertainties is below ${r_{sky}}$. If so, I calculate $v_{sky}$ conventionally (Equation \ref{v_p_determination}). Otherwise, directly using Gaia data would likely be less accurate than using the method of Section \ref{rp_trick}, so my analysis tries it out. To do so, the distances to both stars are first set to their mean i.e.
\begin{eqnarray}
	d_{1,2} \to \frac{d_1 + d_2}{2} \, .
\end{eqnarray}
I then use $r_{sky}$ to obtain a probability distribution for the 3D separation $r_{rel}$ (Equation \ref{x_posterior}). This is sampled using a MC scheme, thereby obtaining the line of sight separation
\begin{eqnarray}
	r_{rel, LOS} \equiv \sqrt{{r_{rel}}^2 - {r_{sky}}^2} \, .
\end{eqnarray}
I then use another random number to decide which star should be further from us. If this is $< 0.5$, star 1 is further away, requiring me to set
\begin{eqnarray}
	d_1 &\to& d_1 \, + \, \frac{r_{rel, LOS}}{2} \, , \\
	d_2 &\to& d_2 \, - \, \frac{r_{rel, LOS}}{2} \, .
\end{eqnarray}
The signs are switched if star 2 needs to be further away than star 1.

I initially focus my analysis on bin 2 (Table \ref{r_sky_edges}), the most relevant for the WBT. The control distribution of the rms $v_{_{line}}$ is shown in Figure \ref{Bin_2_v_line_sky} along with the rms $v_{_{line}}$ of the original data. For comparison, I also show the corresponding quantities if $v_{sky}$ is used instead, including for the case labelled `using $r_{sky}$' where I apply the technique described in Section \ref{rp_trick}. In these cases, the results are divided by $\sqrt{2}$ to allow a fair comparison with the rms $v_{_{line}}$.

Figure \ref{Bin_2_v_line_sky} shows that the observed rms $v_{sky}$ can be adequately explained if $v_{rel} = 0$ for the 22 WBs analysed therein. Consequently, the WBT is likely to prove very difficult using $v_{sky}$. The prospects look much better if using $v_{_{line}}$ because its observed rms value clearly requires $\sigma_{_{1D}}$ to have a non-zero latent value. The WBT also appears feasible if $v_{sky}$ is calculated using the method discussed in Section \ref{rp_trick}.

\begin{figure}
	\centering
	\includegraphics[width = 8.5cm] {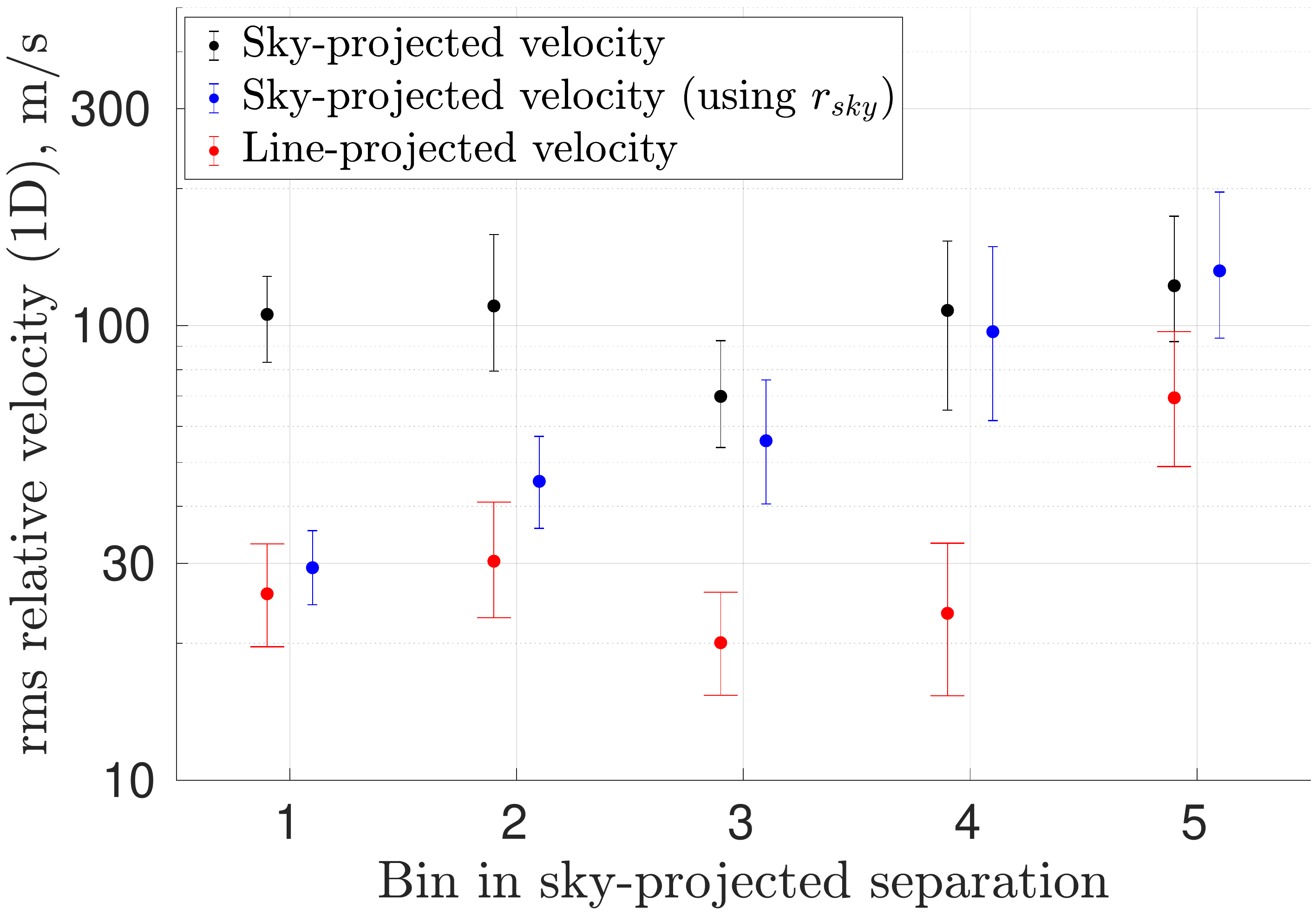}
	\caption{Similar to Figure \ref{Hernandez_v_control}, but now showing the conventional line (red) and sky-projected (black) velocities slightly to the left of my result for $v_{sky}$ calculated using the method of Section \ref{rp_trick}, shown here in blue. The latter method works well for $r_{sky}$ bins 1 and 2, which are the critical bins for the WBT.}
	\label{Hernandez_v_control_rptrick}
\end{figure}

To summarise probability distributions like those shown in Figure \ref{Bin_2_v_line_sky}, I extract the most likely value and 68.3\% confidence interval, equivalent to the central standard deviation of a Gaussian. The most likely value of any quantity $x$ is simply the mode of its probability distribution $P \left( x \right)$, normalised such that $\int P \left( x \right) dx = 1$. To get the confidence interval, I find the value $\alpha$ such that $\int P \left( x \right) dx = 0.683$ if the integral is taken over only that range of $x$ for which $P \left( x \right) > \alpha$. This range is easily determined for unimodal distributions of the sort which arise in this work. Once the appropriate value of $\alpha$ is found, the corresponding range of $x$ defines the 68.3\% confidence interval.

I use Figure \ref{Hernandez_v_control} to show these summary statistics for my control analyses of all $r_{sky}$ bins. As anticipated by \citet{Banik_2018_Centauri}, use of full 3D relative velocities leads to rather large uncertainties. In reality, these may be even larger because I assume that any star with a missing radial velocity has a valid measurement with the same accuracy as for its WB companion.

The use of sky-projected velocities reduces uncertainties somewhat, but Figure \ref{Bin_2_v_line_sky} shows that these are probably still too large in one of the most important $r_{sky}$ bins for the WBT. Uncertainties can be reduced by another factor of ${\approx 4}$ using the line velocity method. In this case, the rms $v_{_{line}}$ would typically be $\la 40$ m/s if its latent value is always 0. Given that $\sigma_{_{1D}}$ must be $\approx 200$ m/s \citep[][figure 7]{Banik_2018_Centauri}, it can be accurately measured using line velocities.

The prospects lie somewhere between these cases if $v_{sky}$ is calculated using the method of Section \ref{rp_trick}. This technique is able to constrain $r_{rel, LOS}$ to within an uncertainty of ${\approx r_{sky}}$. However, $r_{rel, LOS}$ would in any case be known to ${\approx 110}$ kAU if Gaia data are used directly \citep[][section 6.2]{Banik_2018_Centauri}. Thus, my analysis is very likely to switch to a conventional $v_{sky}$ calculation for $r_{sky}$ bins 4 and 5, meaning that no benefits are derived from the method of Section \ref{rp_trick}. It offers modest benefits in bin 3 and substantial benefits in bins 1 and 2 (Figure \ref{Hernandez_v_control_rptrick}). These are the critical bins for the WBT \citep{Pittordis_2018, Banik_2018_Centauri}. Although the uncertainties in these bins are slightly larger than with $v_{line}$ alone, using $v_{sky}$ has the advantage of utilizing two components of $\bm{v}_{rel}$ instead of just one. This could significantly improve the statistical power of the WBT (Section \ref{Effect_on_P_detection}).

\subsection{Inferred velocity dispersions}

My results in Section \ref{Effect_on_P_detection} show that line velocities are expected to follow a roughly Gaussian distribution. Therefore, I repeat my MC trials with an extra Gaussian dispersion of $\sigma_{_{1D}}$ added to each component of $\bm{v}_{rel}$. I then find the proportion of MC trials in which the rms $v_{_{line}}$ of this mock dataset falls within a narrow range around the observed value. This is the relative probability of the particular $\sigma_{_{1D}}$ value used.

As discussed in Section \ref{Measurement_uncertainties}, the result is very small for $\sigma_{_{1D}} = 0$. As $\sigma_{_{1D}}$ is increased, the probability rises up to some maximum before decreasing again. This is because adding a very high $\sigma_{_{1D}}$ causes the rms $v_{_{line}}$ to exceed the observed value in nearly all MC trials.

\begin{figure}
	\centering
	\includegraphics[width = 8.5cm] {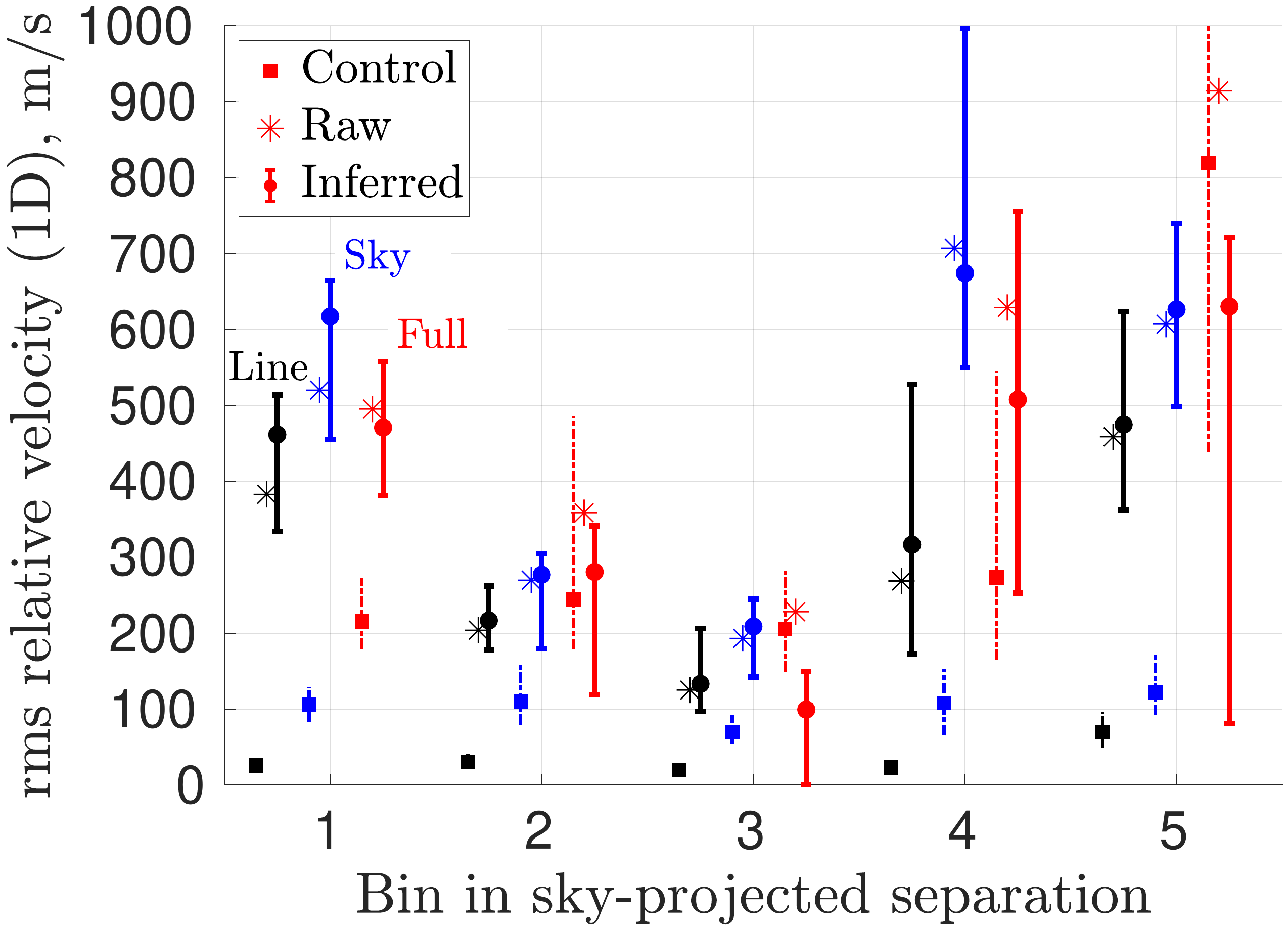}
	\caption{The inferred $\sigma_{_{1D}}$ of WBs in each $r_{sky}$ bin (Table \ref{r_sky_edges}) and their uncertainties (solid circles and error bars). I also show the results of my control analyses where WBs have no relative velocity (solid squares with dashed error bars that are sometimes smaller than the marker). The rms velocity dispersion of the WBs in each bin (* markers) are shown for comparison $-$ these do not require a MC analysis. Results are shown using line velocities (black), sky-projected velocities (blue) and 3D velocities (red). Within each bin, the $x$ co-ordinate is staggered by $\pm \frac{1}{4}$ for clarity. The 3D results are illustrative only as some stars lack radial velocity data (Section \ref{Quality_cuts}).}
	\label{Hernandez_sigma_inference}
\end{figure}

Having obtained an inference on $\sigma_{_{1D}}$, I determine its 68.3\% confidence interval (Section \ref{Measurement_uncertainties}) and show the results in Figure \ref{Hernandez_sigma_inference}. This allows a comparison with the observed rms $v_{_{line}}$ and the results of my control analysis (Section \ref{Measurement_uncertainties}). For bins 2 and 3 which are most relevant to the WBT (Table \ref{r_sky_edges}), $\sigma_{_{1D}}$ is clearly detected. For comparison, I repeat my analyses using the sky-projected and 3D relative velocities, though the results need to be scaled down by factors of $\sqrt{2}$ and $\sqrt{3}$, respectively. Because radial velocities are missing for some stars, the 3D results should be considered illustrative only.

Figure \ref{Hernandez_sigma_inference} shows that using $v_{_{line}}$ and $v_{sky}$ yield rather similar errors. This will change over time because uncertainties in distances are expected to drop slower than those in proper motions (Section \ref{Discussion_tangential_velocity}). Moreover, the small sample size imposes a rather high floor on the uncertainties, even if perfect data were available. It will be interesting to apply the line and sky velocity methods to a larger sample of WBs.

\begin{figure}
	\centering
	\includegraphics[width = 8.5cm] {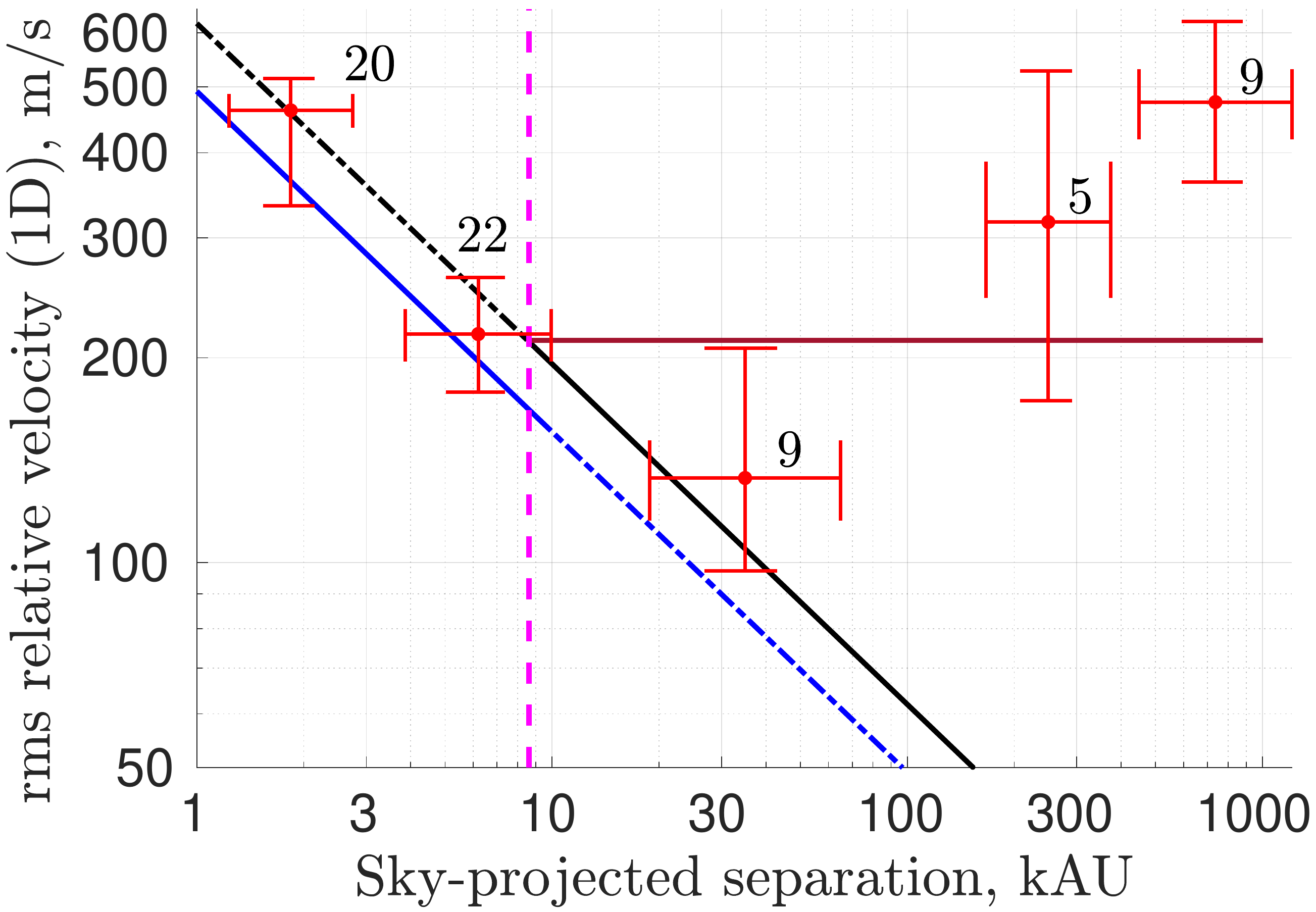}
	\caption{My inferred $\sigma_{_{1D}}$ using line velocities, shown against the mean and dispersion of $r_{sky}$ for the stars in each bin (red points with error bars). The numbers to the top right of each point give the number of systems in that bin after outlier rejection (Section \ref{Quality_cuts}). The lower blue line is the prediction of Newtonian gravity while the upper black line is the asymptotic prediction of MOND, assuming all WBs have a total mass of ${1.5 M_\odot}$ (see text). The pink vertical line is the MOND radius for this mass (Equation \ref{Deep_MOND_limit}). The MOND expectation with the EFE is roughly given by the lower blue line out to the MOND radius and the upper black line beyond it, thus following the solid parts of both lines. The form of the transition is not shown here. Some MOND models lack the EFE, in which case $\sigma_{_{1D}}$ should flatten off at the level indicated by the horizontal dark red line. The predictions shown here are based on \citet[][figure 7]{Banik_2018_Centauri}. As the nearest star to the Sun is 268 kAU away \citep{Kervella_2016}, WBs in bins 4 and 5 are likely to be affected by other stars and are thus unsuitable for the WBT.}
	\label{Hernandez_v_rp_comparison}
\end{figure}

The results in Figure \ref{Hernandez_sigma_inference} allow for a preliminary comparison with theory. For this purpose, I plot my $\sigma_{_{1D}}$ inferences against the mean $r_{sky}$ for the WBs in each $r_{sky}$ bin (Figure \ref{Hernandez_v_rp_comparison}).\footnote{Due to the use of logarithmic axes, I determine the mean and dispersion in $\ln r_{sky}$ and then exponentiate.} Theoretical expectations require knowledge of the WB masses, which I hope to estimate and use in a future analysis. For now, I simply assume that all WBs have a total mass of ${1.5 M_\odot}$, the same assumption made in \citet[][figure 7]{Banik_2018_Centauri} because ${1.5 M_\odot}$ is nearly the mode of the expected Gaia WB mass distribution (see their figure 2). Based on their figure 7, I assume that the Newtonian expectation is 155.7 m/s for $r_{sky} = 20$ kAU while the MOND expectation is 195.9 m/s for conventional versions of it that include the external field effect (EFE). As these are predictions for sky-projected velocities, I scale them down by $\sqrt{2}$ and assume a Keplerian ${r_{sky}}^{-1/2}$ law to obtain results for other $r_{sky}$. This is valid in Newtonian gravity and also in MOND for systems wider than their MOND radius of 8.6 kAU (Equation \ref{Deep_MOND_limit}), since in the Solar neighbourhood such systems are dominated by the EFE such that MOND boosts the Newtonian forces by a fixed factor of ${\approx 1.4}$ \citep[][figure 1]{Banik_2018_Centauri}.

Thus, local WBs within their MOND radius should follow a Keplerian law with the Newtonian normalisation. In Newtonian gravity, the same normalisation should of course remain valid for larger radii. However, in MOND models with the EFE, the normalisation would asymptotically be $\approx 1.2 \times$ higher. Without the EFE (as discussed in their section 7.4), the Keplerian law would no longer apply beyond the MOND radius. Instead, $\sigma_{_{1D}}$ should become independent of $r_{sky}$, reminiscent of flat galactic rotation curves. Based on \citet[][figure 7]{Banik_2018_Centauri}, the asymptotic value should be $\approx 300/\sqrt{2}$ m/s.

These predictions are valid for WBs unaffected by tides from other stars. Given that the nearest star to the Sun is 268 kAU away \citep{Kervella_2016}, systems with $r_{sky} \ga 100$ kAU (bins 4 and 5) are unsuitable for the WBT. Even if such systems are isolated now, it is quite likely they have been disrupted by tides from other stars at some time in the past. The lower orbital velocity and wider separation of such systems makes them particularly vulnerable to tides. As an example, two Sun-like stars orbiting each other in the Solar neighbourhood have a Newtonian Jacobi/tidal radius of 350 kAU \citep[][equation 43]{Jiang_2010}. Considering systems with such wide separations also makes it much more likely for my sample to include ionized systems which have not yet dispersed \citep[][section 8.1]{Banik_2018_Centauri}.

Bearing these expectations in mind, my results in Figure \ref{Hernandez_v_rp_comparison} show that the uncertainties are likely still too large to allow the WBT, at least with the \citet{HERNANDEZ_2018} sample of WBs. Nonetheless, the expected Keplerian decline is clearly evident out to the MOND radius and are suggestive of a further decline beyond it. This implies a mild amount of tension with MOND models that lack an EFE. However, the small sample size and lack of system masses means that one should not draw strong conclusions at this stage.

\section{Testing gravity with line velocities}
\label{Effect_on_P_detection}

In the short term, the WBT will involve $r_{sky}$ rather than the true 3D separation $r_{rel}$ (Section \ref{Sky_projected_separation}). Thus, I follow \citet{Pittordis_2018} and \citet{Banik_2018_Centauri} in defining the scaled relative velocity
\begin{eqnarray}
	\widetilde{v} ~\equiv ~ v_{rel} \div \overbrace{\sqrt{\frac{GM}{r_{sky}}}}^{\text{Newtonian }v_c} .
	\label{v_tilde}
\end{eqnarray}
The sky-projected component of this is $\widetilde{v}_{sky}$ while the line-projected component is $\widetilde{v}_{_{line}}$.

Because $r_{sky}$ measures only part of the 3D $\bm{r}_{rel}$, $\widetilde{v}$ is smaller than what it would be if the full $\bm{r}_{rel}$ were used to calculate it. Thus, $\widetilde{v} < \sqrt{2}$ in Newtonian gravity. The same limit applies to $\widetilde{v}_{sky}$ and $\widetilde{v}_{_{line}}$, though projection effects imply smaller typical values.

\citet[][section 2.2]{Banik_2018_Centauri} derived an analytic estimate for the corresponding upper limit in MOND, which they confirmed using numerical simulations (see their section 5).
\begin{eqnarray}
	\widetilde{v} ~\leq~ \sqrt{2 \, \nu_{_{MW}} \left( 1 + \frac{1}{3} \frac{\partial \, Ln \, \nu_{_{MW}}}{\partial \, Ln \, g_{_{N, MW}}}\right)} ~ .
	\label{v_tilde_limit}
\end{eqnarray}
Here, $\nu_{_{MW}}$ is the MOND enhancement to $g_{_{N, MW}}$, the Newtonian gravity exerted by the rest of the Galaxy on the Solar neighbourhood. Although the Galaxy is a disk, the Sun is sufficiently close to its mid-plane and sufficiently far from its centre that the spherically symmetric MOND interpolating function $\nu$ can be used with negligible loss of accuracy \citep[][section 9.3.1]{Banik_2018_Centauri}. Thus, $g_{_{N, MW}}$ can be inferred from the amplitude of the Galactic rotation curve near the Sun.\footnote{Note that this constrains the product $\nu_{_{MW}} g_{_{N, MW}}$, so analytic or numerical root-finding procedures are required to obtain $\nu_{_{MW}}$.} By combining the latest measurements of the Galactic rotation curve with an interpolating function consistent with the RAR, \citet{Banik_2018_Centauri} showed that the upper limit on $\widetilde{v}$ is expected to be 1.68 in MOND, ${\approx 20\%}$ higher than the Newtonian value.

To quantify the distribution of $\widetilde{v}_{_{line}}$, it is necessary to consider WBs with a range of properties. The semi-major axis probability distribution is carefully chosen such that the distribution of $r_{sky}$ matches observations \citep[][section 3.2]{Banik_2018_Centauri}. Similarly to that work, I assume that observational difficulties will prevent the WBT from using systems with $r_{sky} > 20$ kAU. This is likely conservative as the WB catalogue of \citet{Andrews_2018} maintains a low contamination rate out to 40 kAU. Increasing the upper limit on $r_{sky}$ somewhat improves prospects for the WBT \citep[][figure 5]{Banik_2018_Centauri}. However, the improvement is not dramatic because the frequency of WBs declines $\propto {r_{sky}}^{-1.6}$ \citep{Lepine_2007, Andrews_2017} and the MOND enhancement to gravity is nearly flat beyond 20 kAU \citep[][figure 1]{Banik_2018_Centauri}.

In addition to a range of WB orbit sizes, it is also important to consider a variety of shapes. These are parameterized by the orbital eccentricity $e$ and its generalisation to non-Newtonian gravity theories \citep[][section 4.1]{Pittordis_2018}.\footnote{For the general case that the mutual gravity is not parallel to $\bm{r}_{rel}$, I use the definition in \citet[][section 2.3.1]{Banik_2018_Centauri}.} I assume a linear distribution in $e$.
\begin{eqnarray}
	P \left( e \right) ~=~ 1 \, + \,\gamma \left( e - \frac{1}{2} \right) \, .
	\label{P_e}
\end{eqnarray}
I use $\gamma_{_N}$ to denote the value of $\gamma$ used for Newtonian WB models while $\gamma_{_M}$ is used for MOND models. If the context makes clear which gravity theory is being discussed, then I just use $\gamma$. In both cases, the allowed range of values is between $-2$ and 2.

The distribution of system masses is explained in \citet[][section 3.3]{Banik_2018_Centauri}. Due to the EFE, I also consider systems covering all possible angles between the orbital pole and the direction towards the Galactic Centre (see their section 3.4). The parameter space is explored using a full grid method, making the procedure deterministic.

Motivated by difficulties in correcting observed redshifts for stellar convective motions \citep[][section 2.2]{Kervella_2017}, I previously considered the case where only the sky-projected components of $\bm{v}_{rel}$ are used in the WBT \citep[][section 5]{Banik_2018_Centauri}. Observed redshifts still provide a consistency check on whether a system really is a WB, but the accuracies were assumed to be insufficient for direct use in the WBT. Restricting the WBT in this way roughly doubles the required number of systems to $\approx 300$ \citep[][figure 5]{Banik_2018_Centauri}. This is much less than the $\approx 2000$ WBs identified by \citet{Andrews_2018}, suggesting there is significant scope for prioritising data quality over quantity.

To see how my line velocity method further inflates the number of systems needed for the WBT, I begin by obtaining the $\widetilde{v}_{_{line}}$ distribution $P \left( \widetilde{v}_{_{line}} \right)$ in the different models. In general, a WB system has $\widetilde{v}_{_{line}} = \widetilde{v}_{sky} \left| \sin \phi \right|$ for some angle $\phi$ between its sky-projected relative velocity and systemic proper motion. Thus, a particular value of $\widetilde{v}_{_{line}}$ can arise from any situation where $\widetilde{v}_{sky} \geq \widetilde{v}_{_{line}}$. The probability of doing so depends on $P \left( \widetilde{v}_{sky} \right)$ and the likelihood that $\left| \sin \phi \right| = \widetilde{v}_{_{line}}/\widetilde{v}_{sky}$, which is needed to achieve the correct projection effect. As the distribution of $\phi$ is expected to be uniform, I only need to consider the range $\left(0 - \frac{\rm{\pi}}{2} \right)$. Using standard trigonometric results, I get that
\begin{eqnarray}
	P \left(\widetilde{v}_{_{line}} \right) = \int_{\widetilde{v}_{_{line}}}^\infty \frac{P \left( \widetilde{v}_{sky} \right)}{\sqrt{1 - \left(\frac{\widetilde{v}_{_{line}}}{\widetilde{v}_{sky}}\right)^2}} \, d\widetilde{v}_{sky}
	\label{Convolution_trick}
\end{eqnarray}

\begin{figure}
	\centering
	\includegraphics[width = 8.5cm] {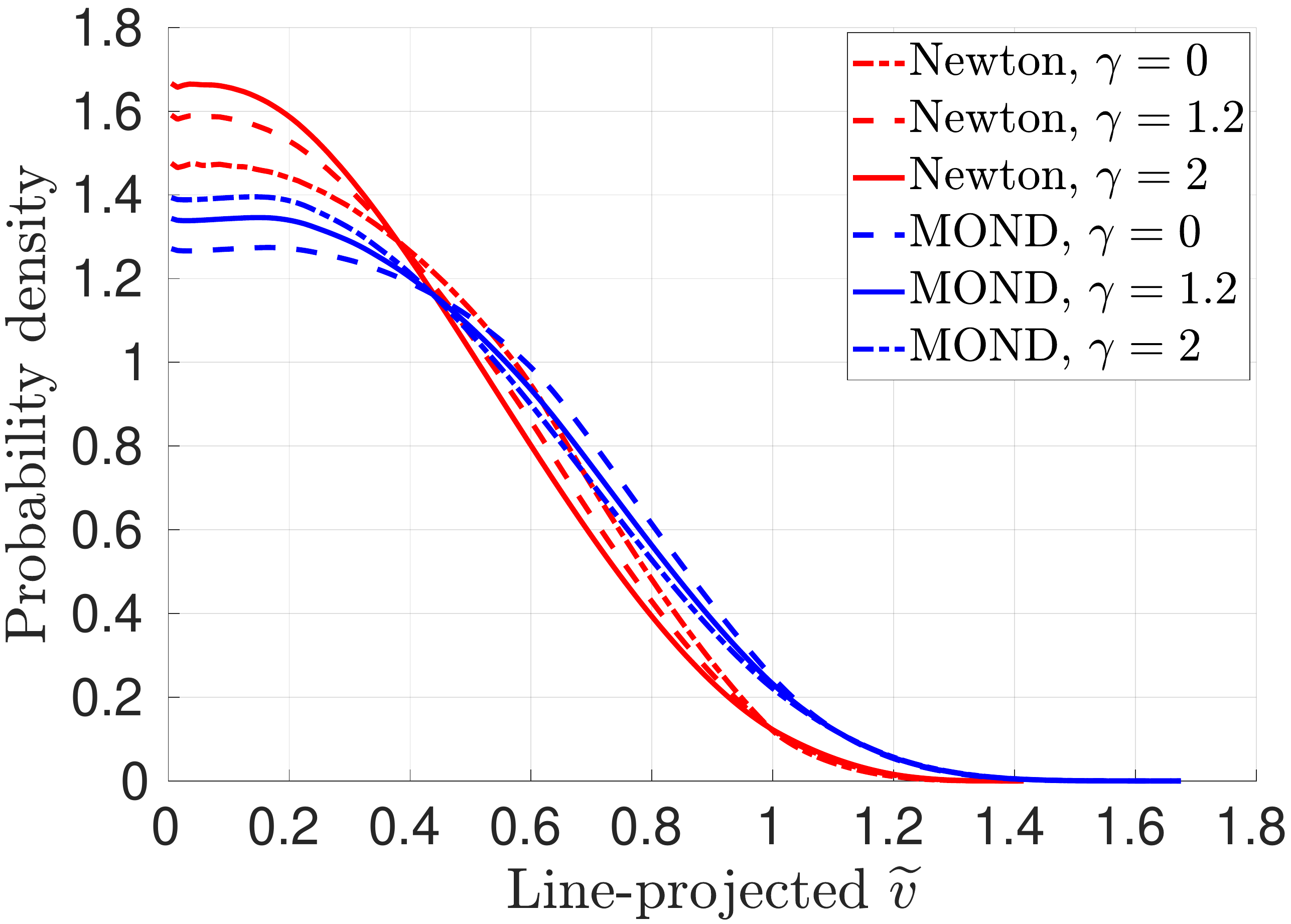}
	\caption{The distribution of $\widetilde{v}_{_{line}}$ for different eccentricity distributions parameterized by $\gamma$ (Equation \ref{P_e}) in Newtonian (red) and Milgromian (blue) dynamics. The results shown here are obtained by applying Equation \ref{Convolution_trick} to $\widetilde{v}_{sky}$ distributions calculated using the methods described in \citet{Banik_2018_Centauri}. Different model parameters are marginalized over using a full grid method, as discussed in their section 3.}
	\label{v_line_comparison_gamma}
\end{figure}

This allows me to take advantage of the $P \left( \widetilde{v}_{sky} \right)$ distributions calculated in \citet{Banik_2018_Centauri}. Some representative examples of $P \left(\widetilde{v}_{_{line}} \right)$ are shown in Figure \ref{v_line_comparison_gamma} for different model assumptions. Within the context of either gravity theory, changing $\gamma$ affects $P \left(\widetilde{v}_{_{line}} \right)$ to a much smaller extent than the difference in $P \left(\widetilde{v}_{_{line}} \right)$ between the different theories. These differences are especially pronounced in the high-velocity tail of the distribution.

Having obtained $P \left(\widetilde{v}_{_{line}} \right)$ for Newtonian and MOND gravity, I use a publicly available method of comparing probability distributions to estimate the probability $P_{detection}$ that these models can be distinguished with accurate data from $N$ systems \citep[][section 4]{Banik_2018_Centauri}. By repeating these `detection probability' calculations for different $N$, I estimate how many systems are required for the WBT and the optimal range in $\left(r_{sky}, \widetilde{v}_{_{line}} \right)$ that astronomers should focus on. For a fixed value of $\gamma_{_M}$, I consider all possible values of $\gamma_{_N}$ in order to find that which minimizes $P_{detection}$. Roughly speaking, this makes the Newtonian $P \left(\widetilde{v}_{_{line}} \right)$ as similar as possible to the MOND $P \left(\widetilde{v}_{_{line}} \right)$. This mimics how future astronomers might try to fit observations of intrinsically Milgromian systems using Newtonian dynamics by adjusting its model parameters.

\begin{figure}
	\centering
	\includegraphics[width = 8.5cm] {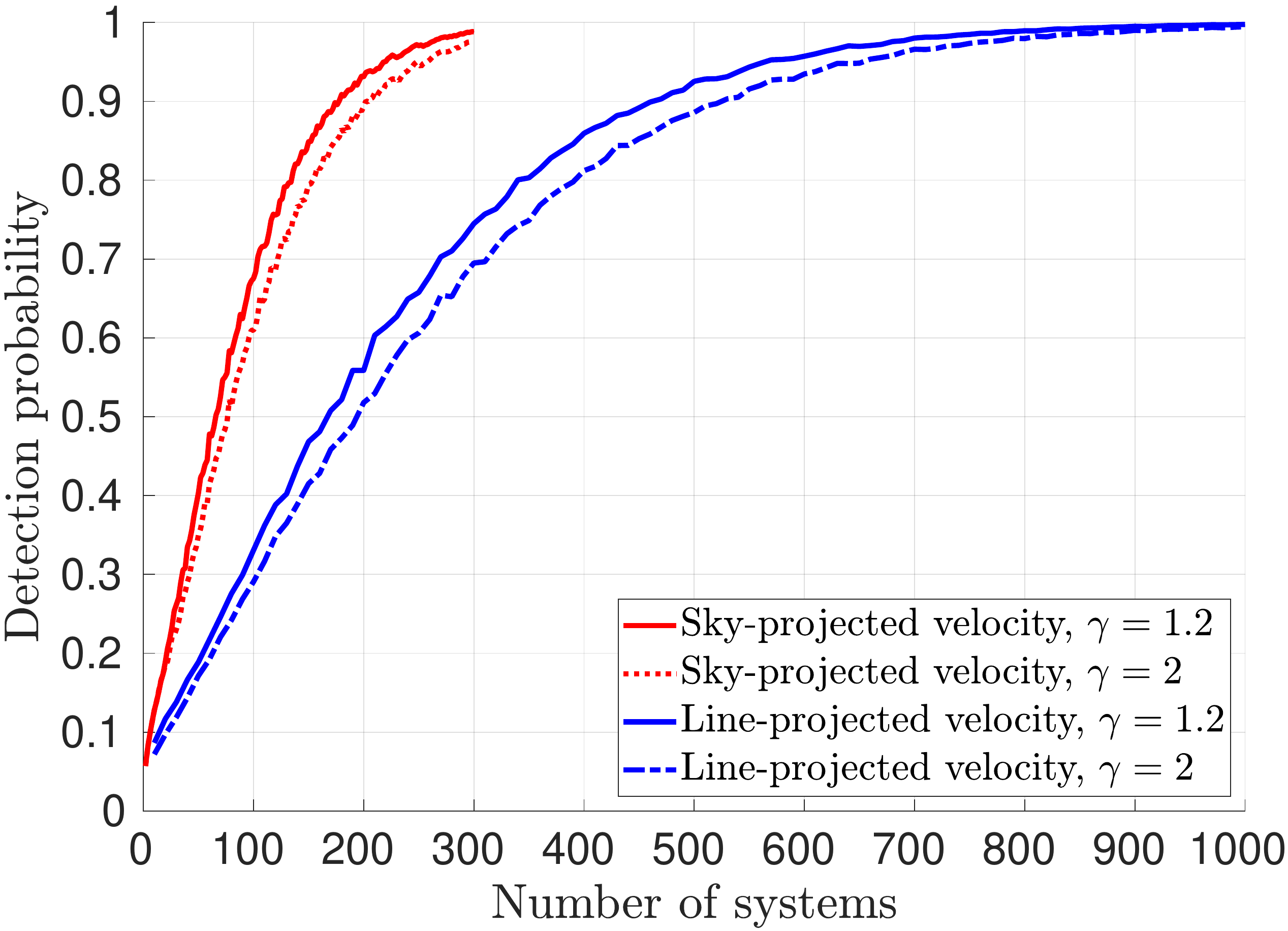}
	\caption{The probability of detecting a significant departure from Newtonian expectations if WB dynamics are governed by MOND and $\gamma_{_M} = 1.2$ (solid lines) or 2 (non-solid lines). Results are shown using line velocities (red) and sky-projected velocities (blue) for different numbers of systems with projected separations of $1-20$ kAU. The values shown here are the minimum attained over all values of $\gamma_{_N}$ \citep[][section 4]{Banik_2018_Centauri}.}
	\label{P_detection_v_line}
\end{figure}

Having obtained $P_{detection}$ in this way, I compare it with similar results based on using $\widetilde{v}_{sky}$ in the WBT (Figure \ref{P_detection_v_line}). As these calculations assume no measurement errors, the line velocity method roughly doubles the $N$ required to reach a fixed $P_{detection}$. This is unsurprising given that $\widetilde{v}_{_{line}}$ is based on only one component of $\bm{v}_{rel}$ whereas $\widetilde{v}_{sky}$ is based on two components.

My calculations yield an a priori estimate of the optimal parameter range for the WBT based on the proportion of systems expected to be in this range under the different gravity models. For my nominal assumption that $\gamma_{_M} = 1.2$, the best $r_{sky}$ range is $3-20$ kAU while the optimal $\widetilde{v}_{_{line}}$ range starts at $0.94 \pm 0.02$ and ends at 1.68, the expected analytic limit (Equation \ref{v_tilde_limit}) and also the maximum value which arises in my MOND models. Out of all WBs with $r_{sky} = 1-20$ kAU, the MOND model predicts that ${3.4 \pm 0.3 \%}$ should fall within this $\left( r_{sky}, \widetilde{v}_{_{line}} \right)$ range. This is nearly triple the Newtonian expectation of ${1.1 \pm 0.2 \%}$ for the `best-fitting' $\gamma_{_N}$ of $\approx -0.5$, the value which minimizes $P_{detection}$. These results are unchanged for $\gamma_{_M} = 2$ apart from the fact that the best $\gamma_{_N}$ rises to $\approx 0.1$. Physically motivated constraints on $\gamma_{_N}$ could improve the prospects for the WBT somewhat, for example if it becomes clear that negative values should not arise.

These results are rather similar to those obtained using $\widetilde{v}_{sky}$ \citep[][section 5]{Banik_2018_Centauri}. The main difference with line velocities is that projection effects significantly reduce the proportion of systems in the optimal parameter range. This makes it more difficult to conduct the WBT, though my results in Figure \ref{P_detection_v_line} indicate that it should still be feasible with ${\approx 1000}$ well-measured systems.

\subsection{MOND without the external field effect}
\label{No_EFE}

\begin{figure}
	\centering
	\includegraphics[width = 8.5cm] {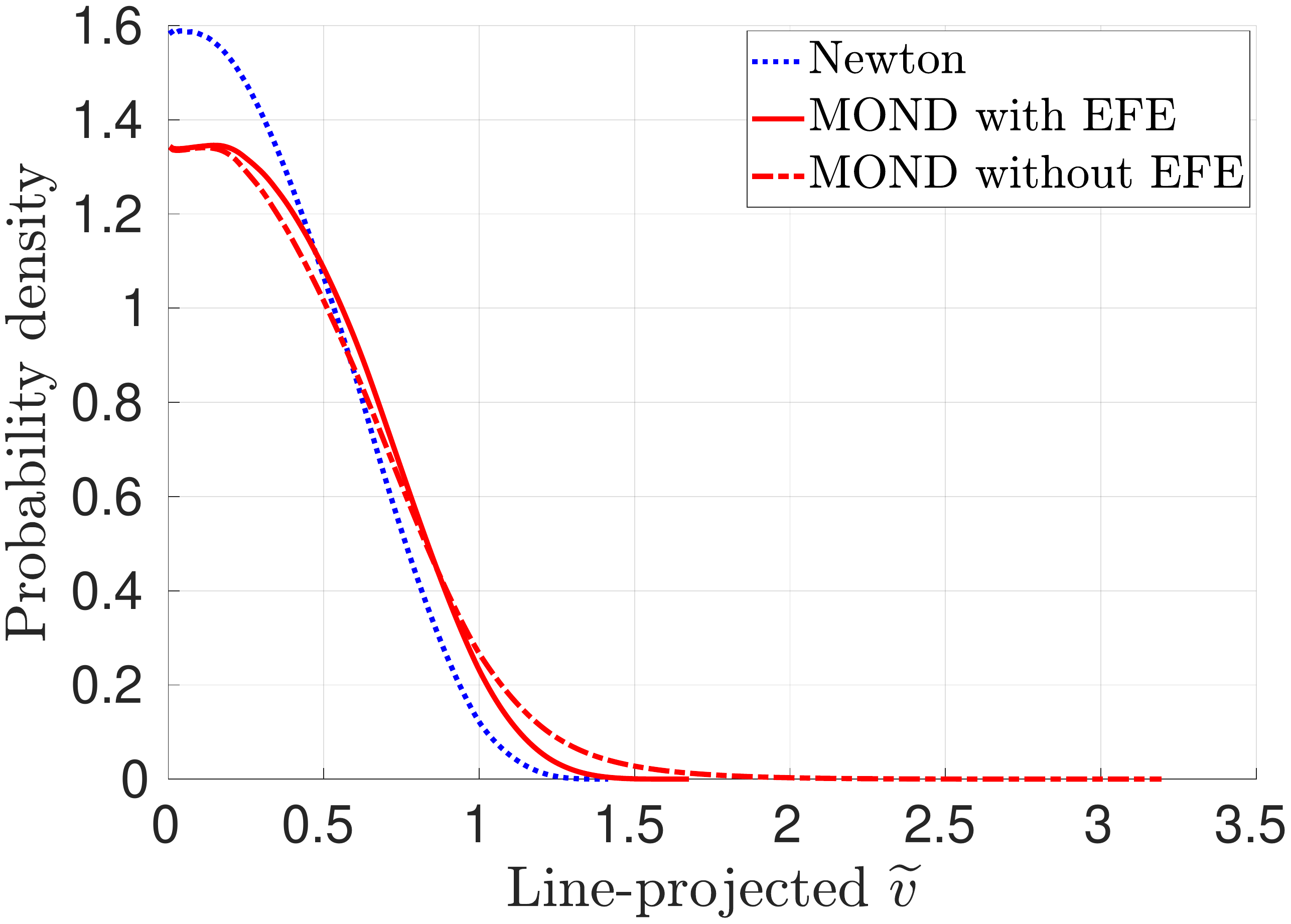}
	\caption{Similar to Figure \ref{v_line_comparison_gamma} for the case $\gamma = 1.2$ \citep{Tokovinin_2016}. In addition to Newtonian dynamics (dotted blue) and conventional MOND (solid red), I also show the case of MOND without the EFE using a dot-dashed red line \citep[][section 7.4]{Banik_2018_Centauri}. In this unconventional scenario, $\widetilde{v}_{_{line}}$ can reach up to 3.2, though it is rarely $\ga 2$.}
	\label{v_line_comparison_no_EFE}
\end{figure}

The EFE is a non-linear effect in MOND which arises directly from its governing equations \citep[][section 2g]{Milgrom_1986}. If a WB system orbits a galaxy with acceleration $\gg a_{_0}$, then the internal dynamics of the WB will be governed by Newtonian gravity regardless of how low its internal accelerations are. This is because the total acceleration enters the governing equations. Therefore, the EFE is not tidal in nature $-$ it arises even if the galaxy exerts a uniform gravitational field across the WB.

So far, the EFE has been directly included in my models \citep[][section 2.1]{Banik_2018_Centauri}. Its section 7.4 discussed the possibility of MOND models without an EFE, as arises in some modified inertia interpretations of MOND \citep{Milgrom_2011}. Despite lacking a self-consistent theory of this type, it is straightforward to repeat my calculations without the EFE as neglecting it greatly simplifies the problem \citep[][equation 13]{Banik_2018_Centauri}.

In Figure \ref{v_line_comparison_no_EFE}, I compare the $P \left(\widetilde{v}_{_{line}} \right)$ distribution in Newtonian gravity against MOND models with and without the EFE. MOND models without the EFE extend out to $\widetilde{v}_{_{line}} = 3.2$ because some systems are much larger than their MOND radius, leading to a large difference compared to more conventional MOND models with the EFE. However, the differences are limited by the rapidly declining $r_{sky}$ distribution of WBs as this implies a similar decline in the distribution of semi-major axes \citep{Andrews_2017}.

\begin{figure}
	\centering
	\includegraphics[width = 8.5cm] {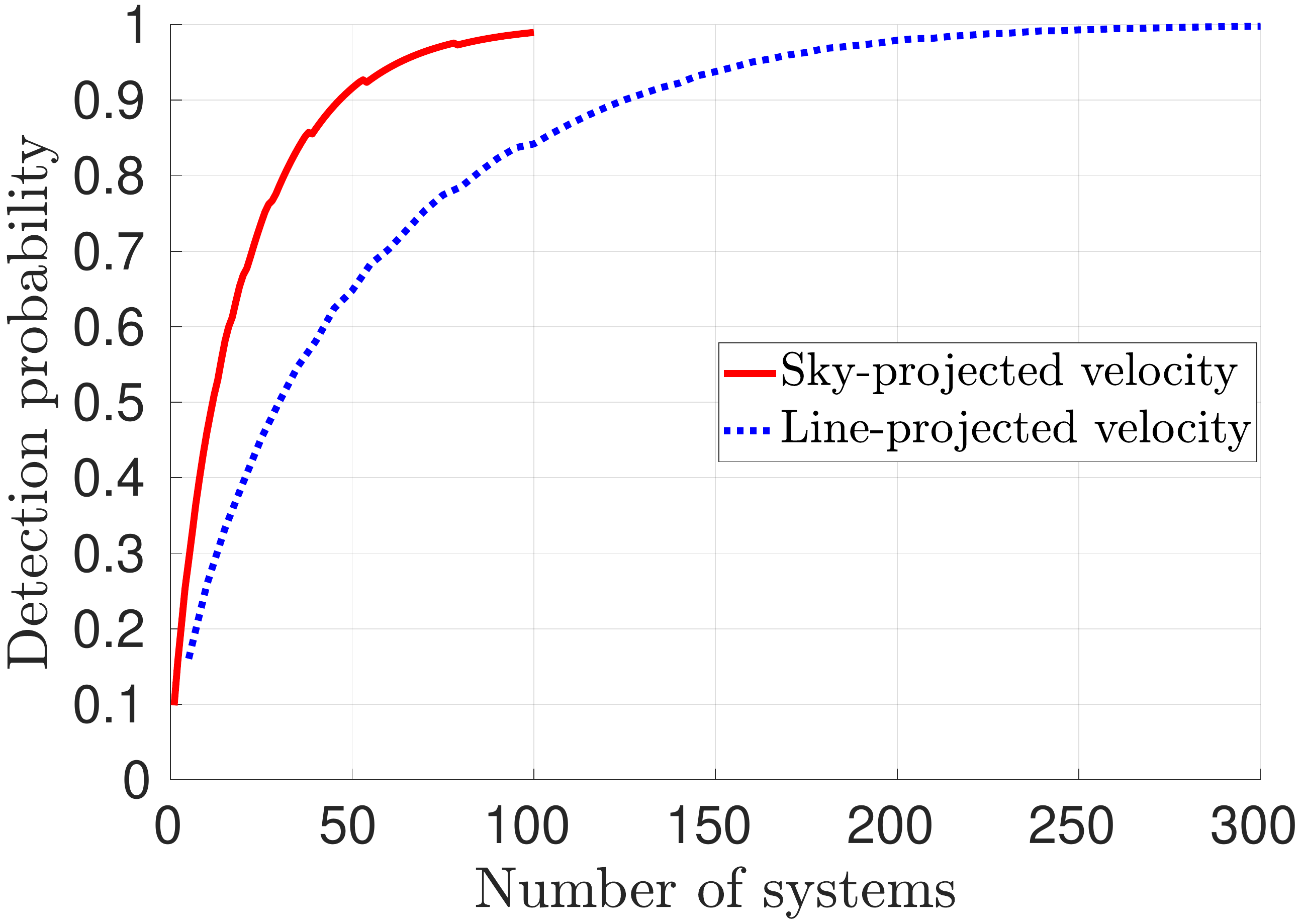}
	\caption{Similar to Figure \ref{P_detection_v_line}, but now showing the distinguishability of Newtonian gravity from MOND without the EFE. These unconventional MOND models predict a more extended $\widetilde{v}_{_{line}}$ distribution (Figure \ref{v_line_comparison_no_EFE}). The larger deviation from Newtonian expectations reduces the number of systems required for the WBT.}
	\label{P_detection_no_EFE_v_line}
\end{figure}

Using my $\widetilde{v}_{sky}$ and $\widetilde{v}_{_{line}}$ distributions for MOND without the EFE, I repeat my $P_{detection}$ calculations and show them in Figure \ref{P_detection_no_EFE_v_line}. These models are much more easily distinguished from Newtonian dynamics (compare with Figure \ref{P_detection_v_line}). Thus, MOND models lacking the EFE will be the first ones to become directly testable using the WBT.

Neglecting the EFE slightly changes the optimal parameter range for the WBT. The best $r_{sky}$ range now becomes $4-20$ kAU while the best $\widetilde{v}_{_{line}}$ range starts at ${0.96 \pm 0.02}$ and extends up to the maximum value of 3.2 reached in my simulations. The best-fitting Newtonian model (${\gamma_{_N} \approx 1.7}$) predicts that ${0.8 \pm 0.2 \%}$ of WBs will fall in this parameter range, much smaller than the ${5.5 \pm 0.2 \%}$ expected in MOND without the EFE. The model predictions would differ even more if accurate data is available for systems with $r_{sky} > 20$ kAU because the lack of an EFE allows MOND to enhance Newtonian accelerations by an unlimited factor.

\section{Discussion}
\label{Discussion}

\subsection{Velocity uncertainties}

My results in Figure \ref{Hernandez_v_control} show that the line velocity technique yields relative velocities with an accuracy of $\approx 30$ m/s. This is very small compared to the expected 1D velocity dispersion of $\approx 150$ m/s in my $r_{sky}$ bin 3 (Figure \ref{Hernandez_v_rp_comparison}). As $v_{_{line}}$ should broadly follow a Gaussian distribution (Figure \ref{v_line_comparison_no_EFE}), measurement errors would increase the width of this distribution by only $\approx \frac{1}{2} \left( \frac{30}{150} \right)^2 = 2\%$, much less than the $\approx 20\%$ difference between orbital velocities in Newtonian and Milgromian dynamics \citep{Banik_2018_Centauri}.\footnote{Similar results would be obtained for bin 2.} Future releases of Gaia data will improve the situation further.

\subsubsection{Tangential velocity}
\label{Discussion_tangential_velocity}

The line velocity method yields such precise results due to its significantly reduced sensitivity to distance uncertainties, which are expected to be larger and decline slower than uncertainties in proper motions. This is because proper motions arise due to the true motion of stars relative to the Sun, which is typically $\sim 30$ km/s \citep{GAIA_2018}. This is about the same as the orbital velocity of Earth around the Sun \citep{Hornsby_1771}, which underlies distance measurements via trigonometric parallax.\footnote{For my discussion, it is sufficient to know the value of 1 AU to within a few percent, which was reliably accomplished in the 1760s using the transits of Venus. More recent determinations confirm the earlier result and further refine it using, amongst other things, spacecraft tracking data and radar reflections off other planets \citep[e.g.][and references therein]{Pitjeva_2009}.} Therefore, the annual parallax of a Solar neighbourhood star is similar to its proper motion over a year. For a fixed astrometric precision, the uncertainty in $v_{rel}$ thus receives similar contributions from distance and proper motion errors after ${\approx 1}$ year of observations.

As observatories such as Gaia \citep{Perryman_2001} collect data over a longer mission duration $T$, proper motion uncertainties are expected to fall as $T^{-3/2}$ because the signal (change in $\widehat{\bm{n}}$) grows linearly with $T$ while measurement errors fall as $\sqrt{T}$ if the frequency of astrometric observations is maintained. However, distances must be inferred from the annual parallax, a cyclical change in $\widehat{\bm{n}}$. Because parallax alone does not cause a long term drift in $\widehat{\bm{n}}$, the distance uncertainty should decrease only as $T^{-1/2}$. Thus, in the long term, the WBT is probably best achieved using line velocities.

If observers achieve better astrometric precision, this would not change the argument because it would improve both distance and proper motion measurements. The only exception is if distance uncertainties somehow `catch up' to those in proper motions, which can be achieved to some extent if $d_{rel}$ is inferred from the rather accurately known $r_{sky}$ (Section \ref{rp_trick}). In the $r_{sky}$ bins most relevant for the WBT, this appears to yield promising results (Figure \ref{Hernandez_v_control_rptrick}), mainly because the method reduces uncertainty in $d_{rel}$ down to ${\approx r_{sky}}$. However, $d_{rel}$ is still rather important to the calculation of $\bm{v}_{rel}$, meaning that its uncertainty would reach a minimum once $d_{rel}$ becomes the dominant source of uncertainty. By contrast, the fractional uncertainty in $v_{line}$ can in principle decrease to the fractional uncertainty in the $d_i$, which is already very small.

The smaller uncertainties resulting from either method suggest that they can be applied to more distant WBs, where a larger uncertainty in the conventional $v_{sky}$ (Equation \ref{v_p_determination}) might make the WB unusable. Similarly, reducing uncertainties might allow the WBT to utilize systems with fainter stars. Assuming such WBs have a lower mass, their reduced MOND radius (Equation \ref{Deep_MOND_limit}) would cause their velocity distribution to differ more significantly between Newtonian and Milgromian dynamics. The statistics are also improved by the use of fainter stars.

These considerations must be set against the simple fact that the line velocity method uses less data per WB, thus inflating the number of WBs needed to distinguish these theories (Section \ref{Effect_on_P_detection}). This deficiency is not present with sky-projected velocities, whose accuracy can be improved substantially if the observed $r_{sky}$ provides a prior on $r_{rel, LOS}$ (Section \ref{rp_trick}). Although the errors are slightly larger than when using line velocities (Figure \ref{Hernandez_v_control_rptrick}), the statistical power of the WBT can be greatly enhanced if it is based on two components of $\bm{v}_{rel}$ rather than just one (Section \ref{Effect_on_P_detection}). Which method will prove more fruitful is unclear at present as this partly depends on the level of contamination, which would affect line velocities more severely (Section \ref{Contamination}). The best option is to try all of them and check if they give consistent results.

\subsubsection{Radial velocity}
\label{Discussion_radial_velocity}

Distance and proper motion uncertainties are undoubtedly the main sources of error when using sky or line velocities. However, the radial velocity does have some impact due to projection effects \citep{Badry_2019}. This can be reduced by revisiting our definition of the direction towards a WB (Equation \ref{n_hat_sys}). This equation is symmetric with respect to its stars. In reality, one of the stars (e.g. star 1) could have a much less well known radial velocity. The line velocity could be made insensitive to this by setting $\widehat{\bm{n}}_{sys} = \widehat{\bm{n}}_1$. In general, instead of using the centre of mass or simply the geometric centre of the stars, one could use the `centre of uncertainty' whereby each star is weighted by the uncertainty in its heliocentric radial velocity. This would yield only modest benefits if the radial velocities are both known to within ${\approx 1}$ km/s as WB angular separations rarely exceed 0.01 radians. If the radial velocity is known for one star but not the other, then it is safe to assume that their radial velocities differ by ${\la 1}$ km/s if the system is a genuine WB. If not, then it would be handled using whatever techniques are used for dealing with contamination, for instance the outlier rejection system used here. Nonetheless, the above-mentioned centre of uncertainty trick is worth trying because it can be used in conjunction with the line velocity method. It is undeniable that the WBT would benefit at least somewhat from exploiting the limited freedom one has in choosing exactly how one defines the direction towards a WB.

In case radial velocities are unavailable for either star, it is still possible to assign each star a value based on its Galactic position and an uncertainty based on the local stellar velocity dispersion. For systems with a small angular separation, this would have only a small effect on the results. However, it is probably not an ideal strategy because the velocity distribution is not Gaussian and the requirement of a small angular separation can severely limit the statistics. Thus, a better method could be to find the relative velocity along the direction $\widehat{\bm{n}}_1 \times \widehat{\bm{n}}_2$, which is completely independent of the stars' radial velocities. Because this is also a measure of relative velocity along a particular line, my results in Section \ref{Effect_on_P_detection} can be used to quantify how much the resulting loss of information inflates the number of systems required for the WBT.

Choosing this statistic comes with the drawback that the relative distance now affects the results, just like with conventional sky-projected velocities. There are two possible solutions to this. Firstly, one can statistically infer the 3D separation from its accurately known sky-projected component (Section \ref{rp_trick}). The resulting uncertainty should be manageable if the angular separation is not too large (Figure \ref{Hernandez_v_control_rptrick}). Alternatively, one could restrict attention to systems where $\widehat{\bm{n}}_1 \times \widehat{\bm{n}}_2$ is nearly parallel to the direction defined by Equation \ref{v_line_direction}, which minimizes the effect of uncertainty in $d_{rel}$. In such systems, it is possible to constrain one component of $\bm{v}_{rel}$ in a way that is almost insensitive to both $d_{rel}$ and the radial velocity of either star.

While my preceding discussion may suggest that only some WBs are suitable for the WBT, one must bear in mind that a lot of the uncertainties scale with the angular separation of the WB. This is smaller for more distant systems, implying that the above-mentioned methods can face difficulties with nearby systems. However, all measured quantities would generally be known more accurately for a WB closer to us, minimizing issues related to uncertainty in $d_{rel}$. The special techniques I discuss for dealing with this uncertainty are both more accurate and more necessary for more distant WBs.

\subsection{Contamination}
\label{Contamination}

In addition to errors in velocity measurements, various other uncertainties would also affect the WBT. Some of these have previously been considered, in particular whether one of the stars in a WB has a close undetected companion as well as its more distant known companion \citep[][section 8.2]{Banik_2018_Centauri}. That work also looked into WBs that were previously ionized by interaction(s) with other stars (see their section 8.1).

If a WB is only marginally ionized, then it will take a long time to disperse. This is quite possible if the ionization is caused by a series of rather weak encounters, as might arise in a star cluster. The end result might be somewhat similar to a moving group of stars \citep{Wielen_1977}. This could add to a background of non-genuine WBs that makes the WBT more difficult. For the particular case of moving groups, the issue could be alleviated somewhat by focusing on systems older than e.g. 1 Gyr. The required stellar age estimates could perhaps be provided by gyrochronology, taking advantage of the increase in stellar rotation periods with age \citep{Barnes_2003}. Very precise ages would not be necessary for this purpose.

By definition, such contamination involves systems whose $\widetilde{v}$ exceeds the limit for bound systems. Due to projection effects, it is possible that $\widetilde{v}_{sky}$ or $\widetilde{v}_{_{line}}$ does not exceed this limit. Statistically, however, it very often will. Thus, observers can estimate the prevalence of contaminating systems by looking at how many WBs have $\widetilde{v}_{_{line}} > 2$, a value almost never exceeded even in versions of MOND without the EFE (Figure \ref{v_line_comparison_no_EFE}).

Nonetheless, contamination would still make the WBT more difficult for the same reason that the brightness of the sky makes it harder to identify a faint astronomical object. To get a feel for how this works, suppose accurate information is available for ${N = 1000}$ systems. My results show that, in the absence of contamination, the WBT will simply be a matter of focusing on a particular range of $\left(r_{sky}, \widetilde{v}_{_{line}}\right)$ and distinguishing between theories which predict 11 vs 34 WBs in this range. Because both numbers are $\ll N$, it is reasonable to assume Poisson statistics. The feasibility of the WBT in this case is just the feasibility of distinguishing Poisson distributions with rates of 11 or 34. Whether these distributions are widely separated can be judged by adding their variances and comparing it to the difference in modes, which correspond to mean values for Poisson distributions. In this case, the means differ by 23 while the difference between random variables following these distributions has an error of $\sqrt{11 + 34} = 6.7$, suggesting a statistically significant exclusion of one or other theory should be possible in the vast majority of cases. This is indeed what my results show (Figure \ref{P_detection_v_line}).

Now suppose that contamination from e.g. moving groups adds an extra 1\% of WBs to this parameter range and that this fact is known based on the distribution of $\widetilde{v}_{_{line}}$ above 2. The gravity theories now predict 21 vs 44 systems in the same parameter range. The difference remains the same but is harder to distinguish, with the uncertainty increasing by a factor of $\sqrt{65/45} \approx 1.2$. To maintain the same statistical significance, ${\approx 1.5 \times}$ as many WBs would therefore be required. Clearly, the WBT would be very challenging if ${\gg 1\%}$ of the systems it is based on fall in the relevant parameter range while not being genuine WBs.

Fortunately, the vast majority of contaminating systems would fall outside this range. For example, if the contamination is uniform in $\widetilde{v}_{_{line}}$ over the range $0-5$ (corresponding to a maximum of 2.1 km/s for two Sun-like stars separated by 10 kAU), then only 14\% of all contaminants would enter the $\widetilde{v}_{_{line}}$ range relevant for the WBT. Thus, the contaminants could comprise up to 7\% of all catalogued WBs with $r_{sky} = \left( 1 - 20 \right)$ kAU.

In reality, an even larger fraction would be tolerable because the contamination can be significantly reduced by using a narrower `aperture' on $\widetilde{v}_{_{line}}$. Figure \ref{v_line_comparison_gamma} shows that MOND predicts almost no systems with $\widetilde{v}_{_{line}} > 1.35$, even though the distribution extends up to 1.68. Thus, the $\widetilde{v}_{_{line}}$ aperture could be narrowed to a width of only 0.4 rather than the 0.7 assumed so far, with only a negligible loss of genuine WBs. Once the prevalence and properties of contaminants are better known, calculations including this information will further optimise the best parameter range to focus on.\footnote{At that stage, it will be difficult to consider these a priori predictions.}

In this context, it should be mentioned that the WB catalogue of \citet{Andrews_2018} has a contamination rate of $\approx 6\%$ while extending out to double the $r_{sky}$ limit of 20 kAU that I assume is observationally accessible. Moreover, the number of WBs they identified greatly exceeds my estimate of how many are required for the WBT (Figure \ref{P_detection_v_line}). This remains true even if my estimate is doubled to account for other sources of uncertainty like contamination.

It must also be borne in mind that the preceding discussion focused on the feasibility of the WBT using only one component of $\bm{v}_{rel}$. At least some information is available regarding the other components, further aiding the WBT.

\section{Conclusions}
\label{Conclusions}

If the anomalous rotation curves of galaxies are caused by a low-acceleration departure from the standard laws of gravity, this will have significant effects on WB systems with separations ${\ga 3}$ kAU. To conclusively perform this WBT and thereby detect or rule out such effects, accurate data is required for systems with separations up to ${\approx 20}$ kAU \citep{Hernandez_2012, Scarpa_2017, Pittordis_2018, Banik_2018_Centauri}. This may already be within reach given that the WB catalogue of \citet{Andrews_2018} extends out to 40 kAU while maintaining a low contamination rate.

At its heart, the WBT requires relative velocities $\bm{v}_{rel}$. Since the test is statistical in nature, it could benefit from considering only the most accurately known component(s) of $\bm{v}_{rel}$. In particular, \citet{Banik_2018_Centauri} considered using only its sky-projected part to minimize the effect of radial velocity uncertainties. \citet[][section 3.2]{Shaya_2011} went a step further by suggesting that only one of the sky-projected velocity components be used. This involves projecting $\bm{v}_{rel}$ onto the direction given by Equation \ref{v_line_direction} and using only this projected quantity in the WBT. The basic principle is to focus on the relative velocity along the direction within the sky plane orthogonal to the systemic proper motion of the WB, thereby minimizing the effect of distance uncertainties. This is because the technique mainly considers the direction of the proper motion vectors rather than their magnitudes.

To demonstrate this line velocity method, I applied it to the WB catalogue of \citet[][table 2]{HERNANDEZ_2018} by conducting MC simulations where measurement errors are included but the stars in each WB have identical latent velocities equal to the mean for the stars in each system. In these control simulations, ${\sigma_{_{1D}} \approx 100}$ m/s when using sky-projected relative velocities but only ${\approx 30}$ m/s using line velocities (Figure \ref{Hernandez_v_control}).

I then performed a preliminary MC analysis of the original \citet{HERNANDEZ_2018} data, finding no evidence of a clear departure from Newtonian expectations at the MOND radii of the systems. My analysis assumed all WBs have a total mass of ${1.5 M_\odot}$ and suffered from a small sample. Even so, the error bars are comparable to the size of the difference between Newtonian and Milgromian expectations. Thus, the WBT should soon become feasible.

To check this, I estimated how many WBs are required to distinguish these theories using the line velocity method. The use of only one component of $\bm{v}_{rel}$ roughly doubles the required number of systems compared to the case where the WBT fully utilizes sky-projected relative velocities. Even so, the WBT should still be feasible with $\approx 1000$ systems (Figure \ref{P_detection_v_line}).

With a longer observing duration $T$, the line velocity method becomes more compelling because it is almost immune to distance uncertainties, which are expected to decrease as $T^{-1/2}$. The method is mainly reliant on proper motions, which should exhibit a more rapid improvement as $T^{-3/2}$. Because distance and proper motion uncertainties should be similar after ${\approx 1}$ year of observations (Section \ref{Discussion_tangential_velocity}), the line velocity technique should be much better after a few years. Its higher accuracy increases the number of usable systems, at least partially offsetting the reduction in how much information is used from each system.

I also discuss how the WBT might be hampered by contamination from unbound systems like moving groups (Section \ref{Contamination}). The effect can be minimized by defining a narrow theoretically motivated range in $\widetilde{v}_{_{line}}$ (Equation \ref{v_tilde_limit}) such that the WBT is best performed by quantifying the proportion of systems in this range \citep{Banik_2018_Centauri}. Using this method, it is likely that the WBT is feasible with the number of WBs reported by \citet{Andrews_2018} if their estimated level of contamination is correct.

Moreover, the WBT will benefit at least somewhat from considering other components of $\bm{v}_{rel}$, even if they do have larger uncertainties. In particular, sky-projected velocities can be made much more accurate if the relative line of sight separation is statistically inferred from the accurately known sky-projected separation (Section \ref{rp_trick}). This technique yields two components of $\bm{v}_{rel}$, one of which could be sacrificed to make the results completely independent of line of sight velocity measurements (Section \ref{Discussion_radial_velocity}). This loss can be avoided if radial velocities are known to within a few km/s, which is not a particularly challenging goal for existing technology but would essentially double the amount of usable data.

Therefore, the next few years promise to bring strong constraints on the behaviour of gravity at the low accelerations typical of galactic outskirts. This will cast a much-needed light on what fundamental new assumptions are required to explain their anomalous behaviour.

\section*{Acknowledgements}

IB is supported by an Alexander von Humboldt postdoctoral fellowship. He wishes to thank the anonymous referee for several very useful suggestions, in particular the use of projected separations to statistically infer true separations. He is grateful to X. Hernandez and R. A. M. Cort{\'e}s for sharing the raw data of their wide binary sample and authorising its distribution. He also thanks H. Zhao for suggesting how to minimize the impact of radial velocity uncertainties, including when these are not known at all. The algorithms were set up using \textsc{matlab}$^\text{\textregistered}$.

\bibliographystyle{mnras}
\bibliography{WBL_bbl} 
\label{lastpage}
\end{document}